\documentclass[submission, Phys]{SciPost}
\usepackage{physics}
\hypersetup{
    colorlinks,
    linkcolor={red!50!black},
    citecolor={blue!50!black},
    urlcolor={blue!80!black}
}

\usepackage[bitstream-charter]{mathdesign}
\DeclareSymbolFont{usualmathcal}{OMS}{cmsy}{m}{n}
\DeclareSymbolFontAlphabet{\mathcal}{usualmathcal}

\begin{document}

	\begin{center}{\Large \textbf{
	Spin-1/2 XX chains with modulated Gamma interaction \\
	}}\end{center}

\begin{center}
	
	M. Abbasi\textsuperscript{1},
	S. Mahdavifar\textsuperscript{1$\star$},
	M. Motamedifar\textsuperscript{2},

\end{center}

\begin{center}

{\bf 1} Department of Physics, University of Guilan, 41335-1914, Rasht, Iran
\\
{\bf 2} Department of Physics, Shahid Bahonar University of Kerman, Kerman, 76169-13439, Iran
\\
${}^\star$ {\small \sf smahdavifar@gmail.com}
\end{center}

\begin{center}
\today
\end{center}

\section*{Abstract}
{\bf
We study the spin-1/2 XX chain with a modulated Gamma interaction (GI), which results from the superposition of uniform and staggered Gamma terms. We diagonalize the Hamiltonian of the model exactly using the Fermionization technique. We then probe the energy gap and identify the gapped and gapless regions. We also examine the staggered chiral, staggered nematic and dimer order parameters to determine the different phases of the ground state phase diagram with their respective long-range orders. Our findings indicate that the model undergoes first-order, second-order, gapless-gapless, and gapped-gapped phase transitions.
}

\vspace{10pt}
\noindent\rule{\textwidth}{1pt}
\tableofcontents\thispagestyle{fancy}
\noindent\rule{\textwidth}{1pt}
\vspace{10pt}

\section{Introduction}

Quantum magnetism investigates how quantum physics influences magnetic systems. Quantum magnets exhibit extraordinary phenomena  such as quantum phase transitions, topological order, and entanglement, while classical magnets do not. Understanding low-dimensional quantum magnets, or magnetic systems with strong interactions in one or two directions, is a significant difficulty in quantum magnetism. These systems can reveal novel phases of matter and events that are both theoretically and empirically accessible  \cite{E1,E2}.

The Heisenberg model \cite{E3-0,E3} is a theoretical framework that characterizes the quantum behavior of magnetic systems, in which atomic spins can engage in diverse forms of interaction. This model is commonly employed to study low-dimensional quantum magnets. The Heisenberg model can be modified to accommodate various lattice geometries and different types of spin interactions, including isotropic or anisotropic, short-range or long-range. These factors influence the many quantum phases and phase transitions that can be observed in the Heisenberg model.

The isotropic Heisenberg model for spin-1/2 chains has no conventional order in the ground state at zero temperature because of the quantum fluctuations \cite{E4-0,E4, E5}. When the spins are perpendicular to a chosen axis and have the same exchange interaction, the model Hamiltonian becomes the spin-1/2 XX Heisenberg chain model \cite{E5-0,E5-0-p1,E5-1,E5-1-p1}. The spin-1/2 isotropic and XX Heisenberg chain models have a gapless excitation spectrum and the spin-spin correlations decay as a power law, which is the main characteristic of an exotic quantum phase called Luttinger liquid phase. The spin liquid phase is a fascinating feature of some low-dimensional magnets that resist a long-range order.

The Kitaev model, which was introduced in $2006$, is a quantum spin model \cite{E5-2}. It describes a system of spin-1/2 particles arranged in a two-dimensional honeycomb lattice in which the direction of the bonds determines the Ising interactions between the spins. The Kitaev model finds application in certain tangible substances as well, including the honeycomb iridates (a class of compounds characterized by a honeycomb lattice configuration and robust spin-orbit coupling) \cite{E5-3} and some of materials such
as $\alpha-RuCl_{3}$, $Na_{2}Co_{2}TeO_{6}$ \cite{E5-3-1, E5-3-2}.  It is established that honeycomb iridates are capable of undergoing quantum phase transitions and new magnetic phases via the GI, which are additional spin interactions beyond the Kitaev model \cite{E5-4}. The GI refers to a form of bond-dependent coupling where the product of two different spin components on neighboring sites is involved, but with different signs or magnitudes for different bonds.

\begin{figure}
	\centerline{\includegraphics[width=0.9\linewidth]{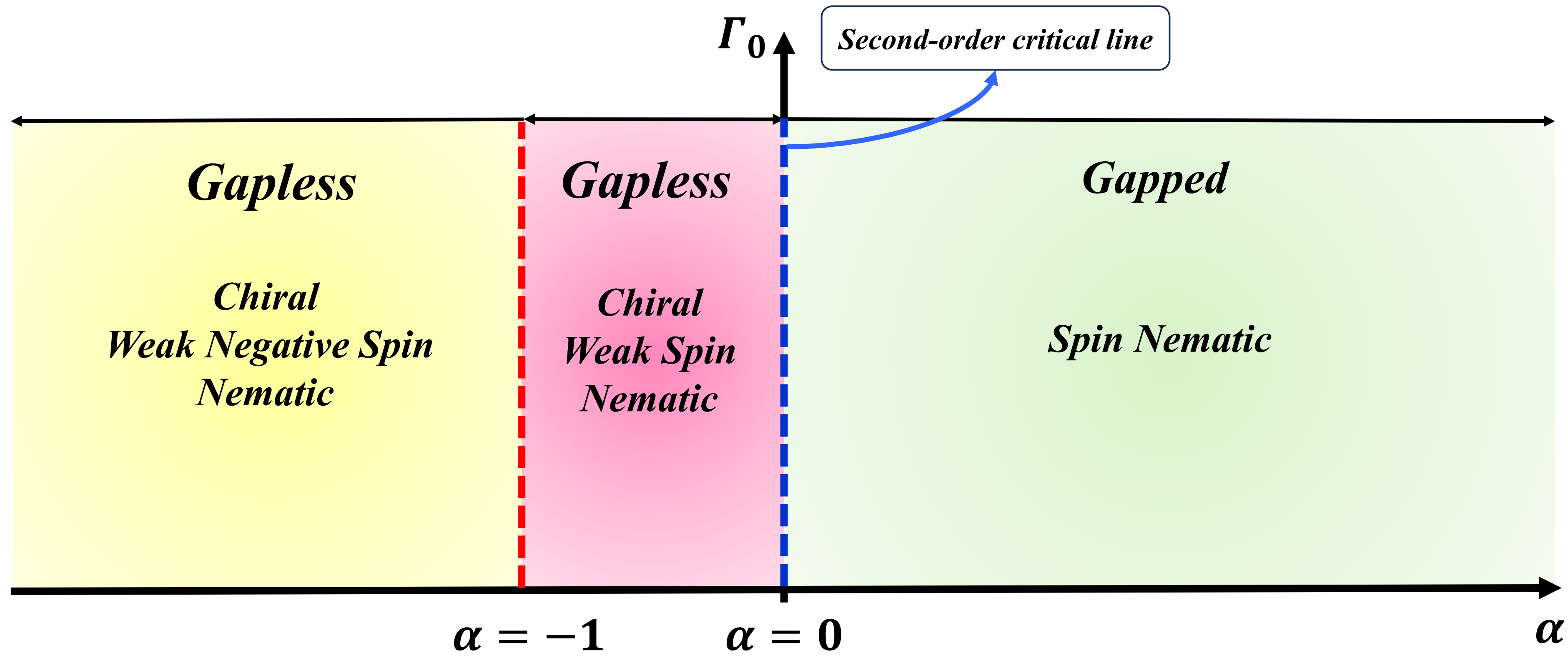}} 
	\centerline{\includegraphics[width=0.9\linewidth]{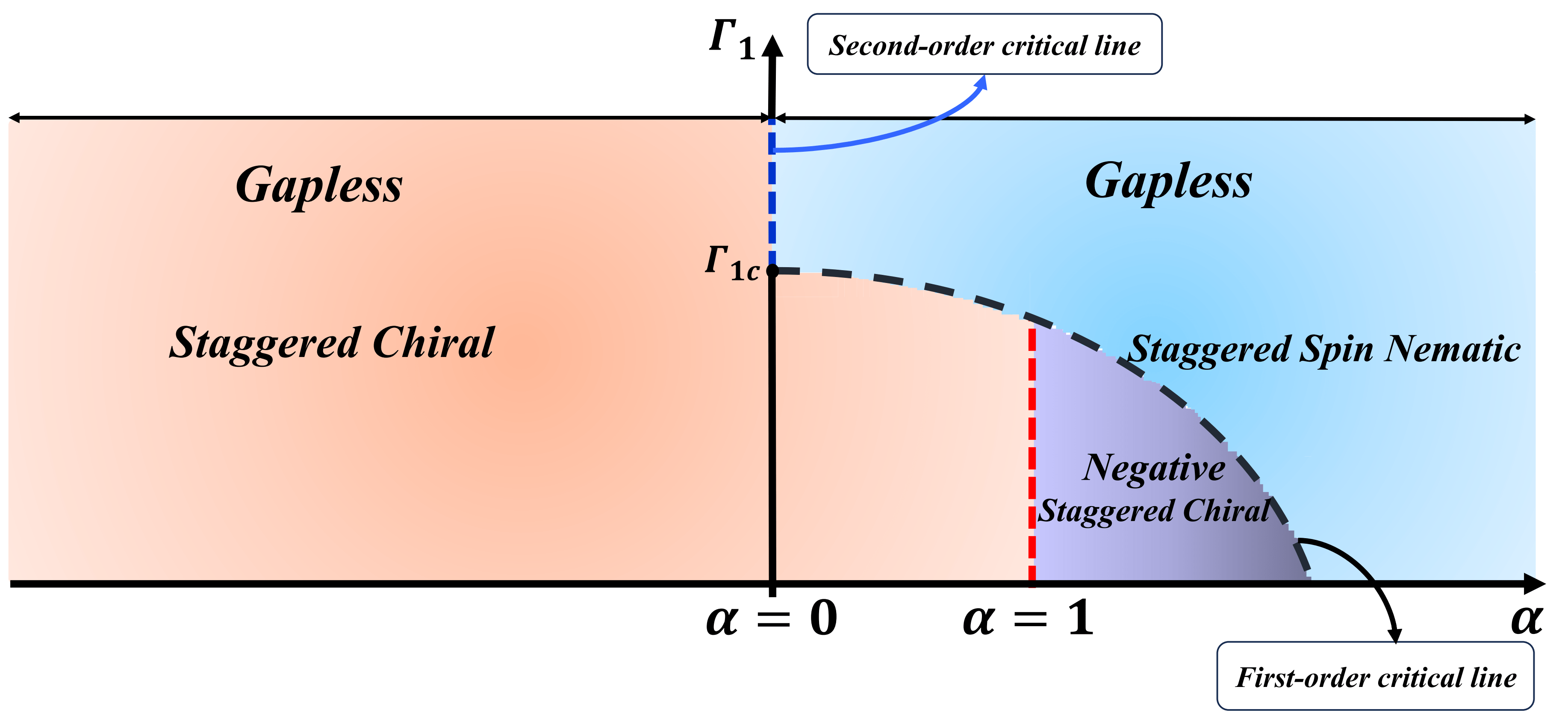}} 
	\centerline{\includegraphics[width=0.9\linewidth]{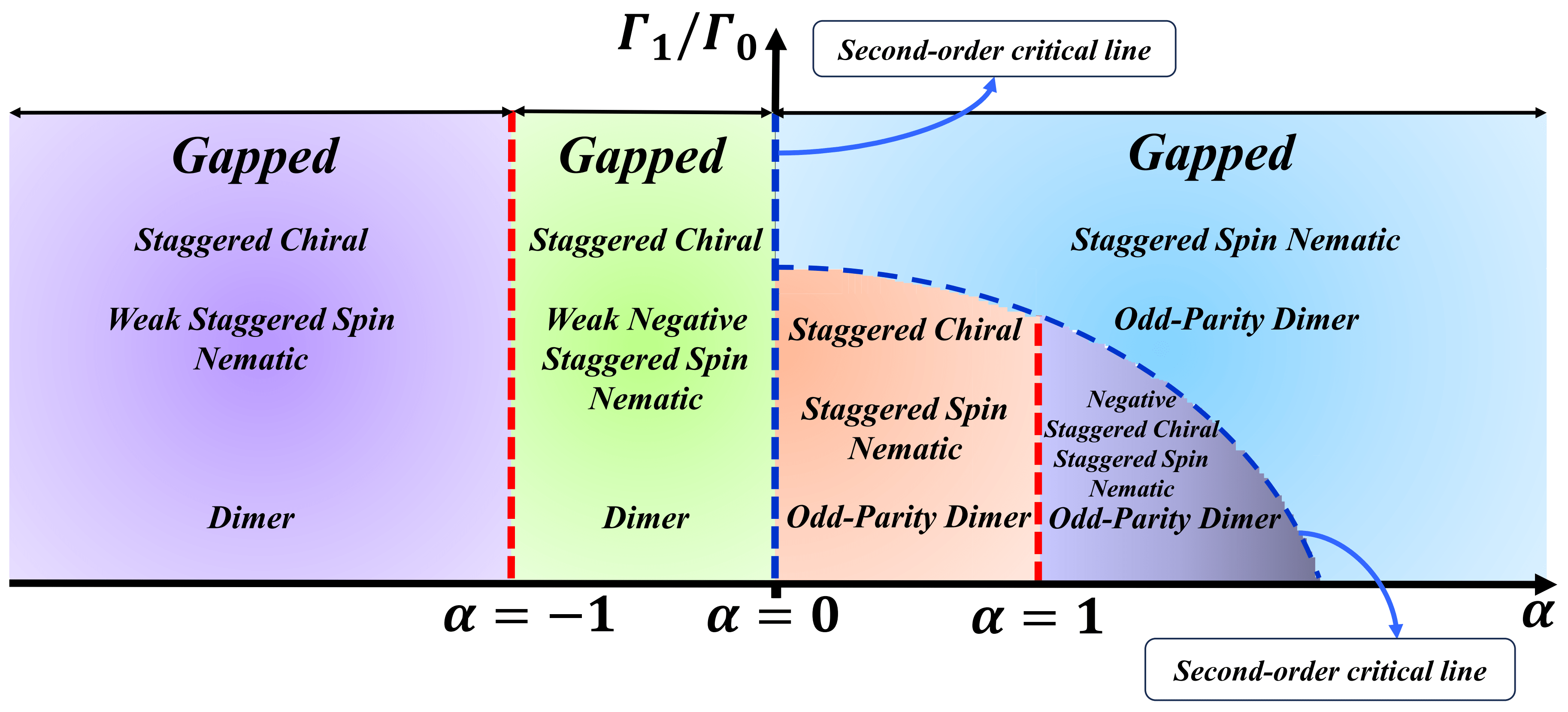}} 
	\caption{Sketch of the ground state phase diagram of the spin-1/2 XX chain with (Top) uniform GI, (Middle) staggered GI and (Bottom) modulated GI.  In the presence of a uniform GI, a critical line is discernible at $\alpha_c=0$, which is unaffected by the value of $\Gamma_0$. There is no spin nematic ordering along the line $\alpha = -1$.  When a staggered GI is applied, a critical line emerges at $\alpha_c=0$ for $\Gamma_1>\Gamma_{1c}$. However, for $\Gamma_1 < \Gamma_{1c}$, a critical line emerges, varying with $\Gamma_1$ as $\alpha_c(\Gamma_1)$. Along the line $\alpha = 1$, no staggered chiral ordering is observed. Lastly, with modulated GI, three critical lines are observed.
	}
	\label{Fig0-}
\end{figure}

Several researches have primarily concentrated on examining the GI in spin-1/2 chain models \cite{E5-5, E5-6, E5-7, E5-8, E5-9-1}. The Gamma model can be implemented using a quantum simulator that utilizes trapped atoms and photons. This implementation yields an effective Hamiltonian \cite{E5-5} for which has been proposed a strategy that employed two types of atoms, one for the spin degree of freedom and another for the photon-mediated interaction. In this reference, the atoms were restricted within an optical lattice and connected to a cavity mode which could function  as a reservoir of photons. As a result, it was obtained an  effective Hamiltonian for the system, which takes the shape of the XY-Gamma model, incorporating additional terms. By manipulating the characteristics of the cavity, such as the frequency, decay rate, and driving field, one can modify the effective Hamiltonian. Moreover, it was illustrated that the effective Hamiltonian could exhibit different regimes, such as the weak-coupling, strong-coupling and resonant regimes, depending on the ratio of the cavity frequency to the atomic transition frequency. Furthermore, it was shown that the effective Hamiltonian could have a variety of interactions, including the Heisenberg, Kitaev, and Gamma couplings. These interactions are contingent upon the detuning of the cavity frequency and the driving field frequency. Ultimately, it was also extracted the phase diagram of the spin-1/2 XY model with GI in its ground state which showed three separate regions: gapless, gapped, and critical. During the gapless phase, the system exhibited an incommensurate spiral order, which was defined by the vector-chiral correlations. On the other hand, during the gapped phase, the system exhibited a topological order, which is defined by the string order parameter. In another research (Ref.~ \cite{E5-7}), the ground phase diagram of the bond-alternating spin-$\frac{1}{2}$ $K$-$\Gamma$ chain was analyzed, revealing that it was primarily characterized by two dominant phases: the even-Haldane phase and the odd-Haldane phase. The even-Haldane phase had no topological properties, but the odd-Haldane phase was a phase with protected symmetry and possessed topological properties. In close proximity to the antiferromagnetic Kitaev limit, there existed two phases with energy gaps, which could be distinguished by their unique nonlocal string correlators.

We study the spin-1/2 Heisenberg XX chain model with a more general spatially modulated GI. The GI can be modulated by changing the bond direction or the spin direction along a chain of spins. This can create a modulation pattern of the GI, which can affect the quantum phases and phase transitions of the system. The GI breaks some symmetries of the system, such as the time-reversal symmetry, the inversion symmetry, and the $U(1)$ symmetry of the XX Heisenberg model. The time-reversal symmetry is the symmetry under the reversal of the direction of time. The inversion symmetry is the symmetry under the inversion of the spatial coordinates. The GI can induce a spin chirality, which is a measure of the twist of the spins around the chain. The spin chirality breaks the time-reversal symmetry, as it changes sign when the time direction is reversed. The modulation pattern of the GI breaks the inversion symmetry, as it changes the spatial parity of the system. In this paper, we show that the breaking of these symmetries can lead to novel phenomena in the quantum phases and phase transitions of the spin-1/2 XX chains. In particular, the ground state phase diagram is illustrated in Figure $1$.   As shown in Fig.~\ref{Fig0-} (a), for the case of uniform GI, three distinct phases are identified: the spin nematic phase and two composite phases. As illustrated in Fig.~\ref{Fig0-} (b), the application of staggered GI reveals tow critical lines. These lines delineate the three gapless phases: staggered chiral, staggered nematic, and negative staggered phases.  Lastly, with the introduction of mutual GI interaction as is seen in Fig.~\ref{Fig0-} (c),  critical lines are discernible, with each ground state phase manifesting as a composite gapped phase.

Therefore, in order to address the aforementioned eventualities in further detail, we organized this study as follows: In the next section, we introduce the model and employ the fermionization approach to derive the system's spectrum. In Section III, we present our findings regarding the ground state phases and critical points. Finally, in Section IV, we provide our conclusions and a summary of the results.

\section{Model}

The Hamiltonian of the 1D spin-1/2 XX  model with modulated GI is denoted as 
\begin{eqnarray}\label{eq1}
	{\cal H} &=&J\sum\limits_{n = 1}^N {\left[ {S _n^x S _{n + 1}^x +  S _n^y S _{n + 1}^y} \right]} - \sum\limits_{n = 1}^N (\Gamma_0+(-1)^{n} \Gamma_1)(S _n^x S _{n + 1}^y+\alpha S _n^y S _{n + 1}^x)~,
\end{eqnarray}
where $S _n$ represents the spin operator on the $n$-th site.  The symbol $J>0$ represents the antiferromagnetic exchange coupling. The amplitudes of off-diagonal exchange interactions are characterized by $\Gamma_0$ and $\Gamma_1$. In this equation, $\alpha$ symbolizes  the relative coefficient between different terms of off-diagonal exchange couplings.   The variable $N$ represents the size of the system, which corresponds to the number of spins. We are assuming the periodic boundary condition, denoted as $S _{n+N}^\mu=S _n^\mu $, where $\mu$ might take the directions of $x$, $y$, or $z$. 

The Hamiltonian can be exactly diagonalized using the fermionization approach. Let us first rewrite Hamiltonian in an conventional form as
\begin{eqnarray}\label{eq1}
	{\cal H} &=&J\sum\limits_{n = 1}^{N/2} {\left[ {S _{2n-1}^x S _{2n}^x +  S _{2n-1}^y S _{2n}^y} \right]} -\Gamma_{-}\sum\limits_{n = 1}^{N/2} (S _{2n-1}^x S _{2n}^y+\alpha S _{2n-1}^y S _{2n}^x)\nonumber \\
	&+&J\sum\limits_{n = 1}^{N/2} {\left[ {S _{2n}^x S _{2n+1}^x +  S _{2n}^y S _{2n+1}^y} \right]} -\Gamma_{+}\sum\limits_{n = 1}^{N/2} (S _{2n}^x S _{2n+1}^y+\alpha S _{2n}^y S _{2n+1}^x)~,
\end{eqnarray}
where $\Gamma_{\pm}=\Gamma_{0} \pm \Gamma_{1}$. Then applying the Jordan-Wigner transformation as \cite{E5-12-0,E5-12,E5-12-p1}
\begin{eqnarray}
	S^{+}_{2n-1}&=&a_{n}^{\dag}e^{i\pi (\sum^{n-1}_{l=1}(a^{\dag}_{l}a_{l}+b^{\dag}_{l}b_{l}))},\nonumber\\
	S^{z}_{2n-1}&=&a_{n}^{\dag}a_{n}-\frac{1}{2},\nonumber\\
	S^{+}_{2n}&=&b_{n}^{\dag}e^{i\pi (\sum^{n}_{l=1}a^{\dag}_{l}a_{l}+\sum^{n-1}_{l=1} b^{\dag}_{l}b_{l})},\nonumber\\
	S^{z}_{2n}&=&b_{n}^{\dag}b_{n}-\frac{1}{2},
	\label{eq17}
\end{eqnarray}
where, $a_n^\dagger$, $a_n$, $b_n^\dagger$, $b_n$, are the fermionic operators, the fermionized form of the Hamiltonian is obtained as 
\begin{eqnarray}\label{eq1}
	{\cal H} &=&\frac{J}{2}\sum\limits_{n = 1}^{N/2} ( a_n^\dagger b_n+b_n^\dagger a_{n+1}+H. c.)\nonumber \\
	&+&\frac{i \Gamma_{-}}{4}\sum\limits_{n = 1}^{N/2} [(1+\alpha) (a_n^\dagger b_n^\dagger- H. c.) -(1-\alpha)(a_n^\dagger b_n-H. c.)]\nonumber \\
	&+&\frac{i \Gamma_{+}}{4}\sum\limits_{n = 1}^{N/2} [(1+\alpha) (b_n^\dagger a_{n+1}^\dagger- H. c.) -(1-\alpha)(b_n^\dagger a_{n+1}-H. c.)].\nonumber 
	\\
\end{eqnarray}
Next, implementing  following Fourier transformations 
\begin{eqnarray}\label{eq1}
	{a_n} &=& \frac{1}{\sqrt{N/2}}  \sum_k e^{ - ikn} {a_k}\nonumber \\
	{b_n} &=& \frac{1}{\sqrt{N/2}}  \sum_k e^{ - ikn} {b_k},
\end{eqnarray}
the Hamiltonian transforms into a sum of commuting Hamiltonians, $H_{k}$, each describing a different $k$ mode,
\begin{eqnarray}\label{eq4}
	{\cal H} &=& \sum\limits_{k>0} H_k \nonumber \\
	&=& \sum\limits_{k>0} [{\cal A}_k (a_k^\dag b_{k}+a_{-k}^\dag b_{-k})+H. c.] +  \sum\limits_{k>0} [{\cal B}_k (a_k^\dag b_{k}-a_{-k}^\dag b_{-k})+H. c.] \nonumber \\
	&+& \sum\limits_{k>0} [{\cal C}_k (a_k^\dag b_{-k}^\dag+ a_{-k}^\dag b_{k}^\dag)+H. c.] + \sum\limits_{k>0} {\cal D}_k (a_k^\dag b_{-k}^\dag- a_{-k}^\dag b_{k}^\dag+H. c.), 
\end{eqnarray}
where 
\begin{eqnarray}\label{eq4}
	{\cal A}_k  &=& \frac{1}{2} J(1+\cos(k)+\frac{i}{4} (1-\alpha) (\Gamma_{-}-\Gamma_{+} \cos(k))\nonumber \\
	{\cal B}_k  &=& [\frac{i}{2} J+\frac{1}{4} \Gamma_{+} (1-\alpha)] \sin (k)\nonumber \\
	{\cal C}_k  &=&  i\frac{1}{4} (1+\alpha) (-\Gamma_{-} + \Gamma_{+} \cos (k))\nonumber \\
	{\cal D}_k  &=&  -\frac{1}{4}(1+\alpha)\Gamma_{+} \sin (k).
\end{eqnarray}
We now can obtain the spectrum of the model by diagonalizing each Hamiltonian mode, $H_{k}$, independently. First, the basis vectors of the sixteen dimensional Hilbert space of $H_{k}$ are chosen as:
\begin{eqnarray}\label{eq4}
	| \phi_{1,k} \rangle &=& |0\rangle~;~| \phi_{2,k} \rangle = a_{k}^{\dag} a_{-k}^{\dag} |0\rangle ~;~| \phi_{3,k} \rangle = a_{k}^{\dag} b_{-k}^{\dag} |0\rangle,  \nonumber\\
	| \phi_{4,k} \rangle&=&  a_{-k}^{\dag} b_{k}^{\dag} |0 \rangle ~;~| \phi_{5,k} \rangle= b_{k}^{\dag} b_{-k}^{\dag} |0 \rangle~;~| \phi_{6,k} \rangle= a_{k}^{\dag} b_{k}^{\dag} |0 \rangle, \nonumber\\
	| \phi_{7,k} \rangle&=&  a_{-k}^{\dag} b_{-k}^{\dag} |0 \rangle ~;~| \phi_{8,k} \rangle= a_{k}^{\dag}a_{-k}^{\dag} b_{k}^{\dag} b_{-k}^{\dag} |0 \rangle, \nonumber \\
	| \phi_{9,k} \rangle&=& a_{k}^{\dag}a_{-k}^{\dag}  b_{k}^{\dag} |0 \rangle~;~ | \phi_{10,k} \rangle= a_{k}^{\dag}a_{-k}^{\dag}  b_{-k}^{\dag} |0 \rangle, \nonumber \\
	| \phi_{11,k} \rangle&=& a_{k}^{\dag}b_{k}^{\dag}  b_{-k}^{\dag} |0 \rangle~;~ | \phi_{12,k} \rangle= a_{-k}^{\dag}b_{k}^{\dag}  b_{-k}^{\dag} |0 \rangle,\nonumber \\
	| \phi_{13,k} \rangle&=& a_{k}^{\dag}|0 \rangle~;~ | \phi_{14,k} \rangle= a_{-k}^{\dag} |0 \rangle~;~ | \phi_{15,k} \rangle= b_{k}^{\dag} |0 \rangle, \nonumber \\
	| \phi_{16,k} \rangle&=& b_{-k}^{\dag} |0 \rangle,
\end{eqnarray}
where $|0\rangle$ denotes the vacuum state of $H_h$. Then, by contracting the matrix form of $H_k$ and diagonalizing it the eigenvalues and eigenstates are obtained. The comparison of these eigenvalues allows us to find the ground state of $H_k$ for every momentum in the region, $k>0$. 

\section{Results}

A notable feature occurs on the special value $\alpha=-1$ where the off-diagonal Gamma interaction becomes the spatially modulated Dzyaloshinskii–Moriya interaction \cite{E5-9, E5-10, E5-11}. In absence of the staggered part, $\Gamma_1=0$, the gapless chiral ordering induces by uniform $\Gamma_0$. As soon as the staggered part is added, a gap emerges in the spectrum and system goes into a composite phase characterized by the coexistence of long-range ordered alternating dimerization and the spin chirality patterns.  In the case of uniform Gamma interaction, $\Gamma_1=0$, a quantum phase transition will happen at the critical $\alpha=0$  \cite{E5-5,E5-6}. The same picture is seen for the case $\Gamma_0=0$.

We start our study with the energy gap \cite{R1-1,R1-2}. The energy gap is defined as the difference between the ground state and the first excited state of a quantum system. It is a measure of the quantum fluctuations and correlations that prevent the system from having static long-range order. The energy gap depends on various factors, such as the strength of the interactions, the external magnetic field, and the presence of disorder or dimerization. Moreover, the energy gap in spin-1/2 chains can be probed experimentally by various techniques, such as inelastic neutron scattering, nuclear magnetic resonance, and optical spectroscopy \cite{R1-4,R1-5}.



\begin{figure}
	\centerline{\includegraphics[width=0.55\linewidth]{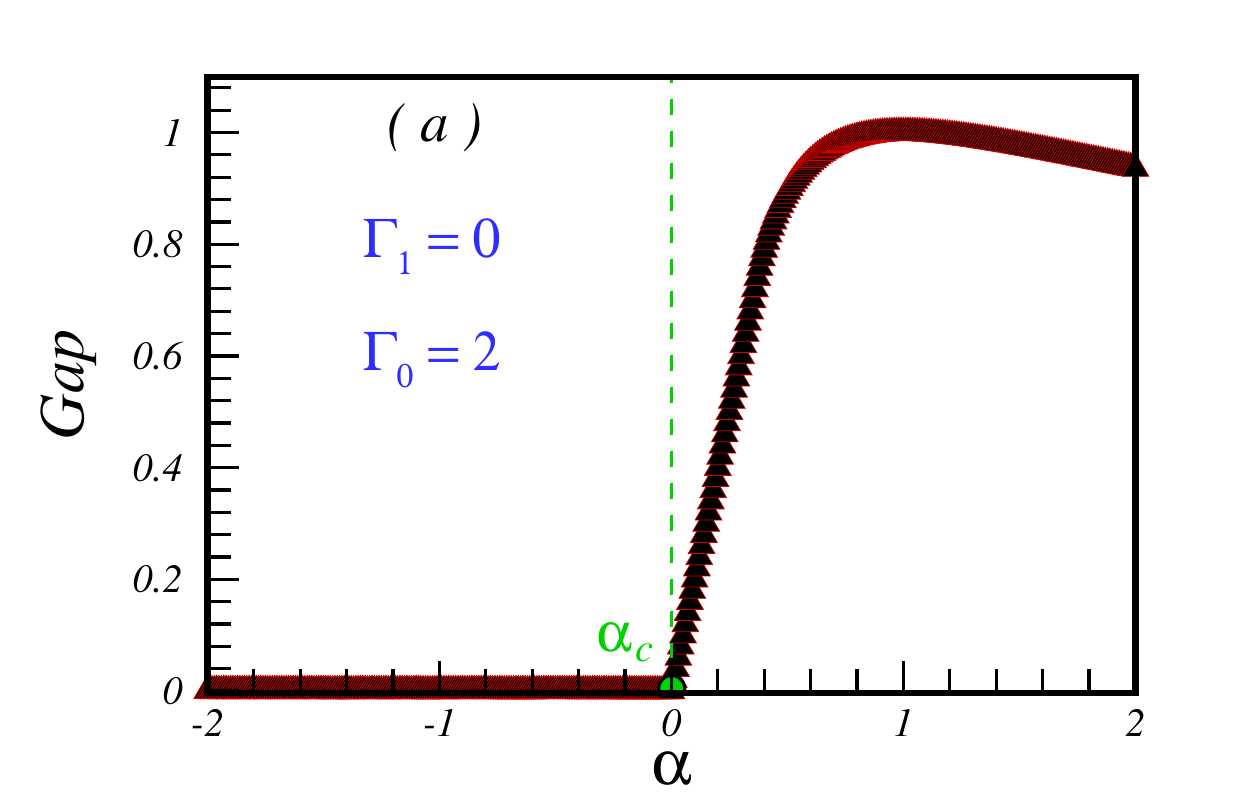}\includegraphics[width=0.55\linewidth]{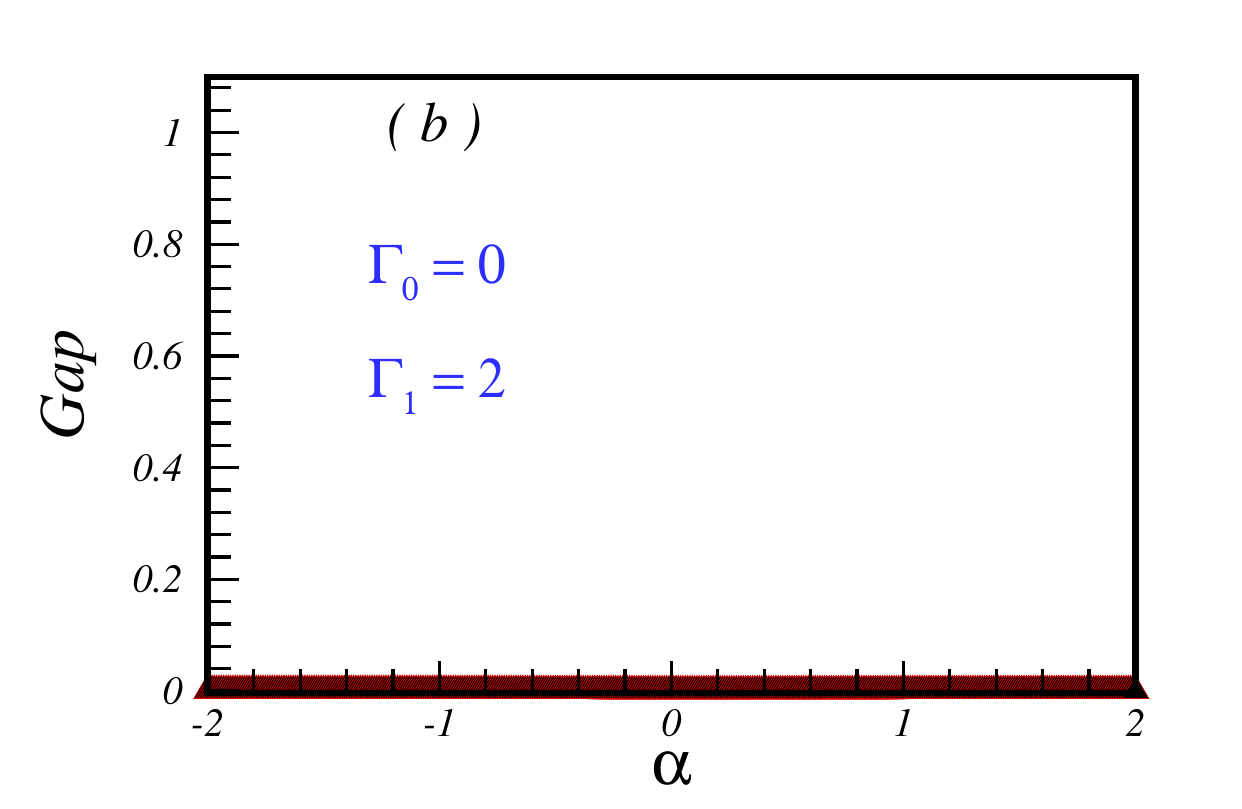}}
	\centerline{\includegraphics[width=0.55\linewidth]{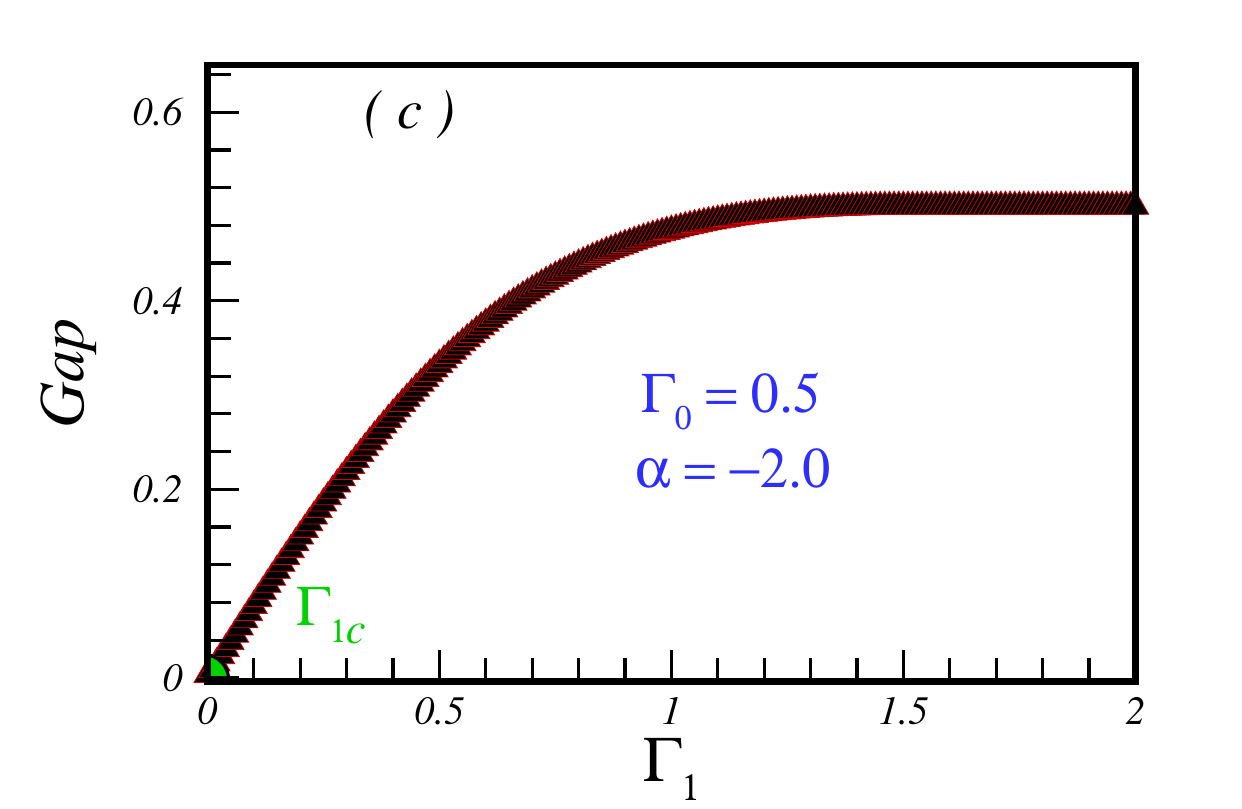}\includegraphics[width=0.55\linewidth]{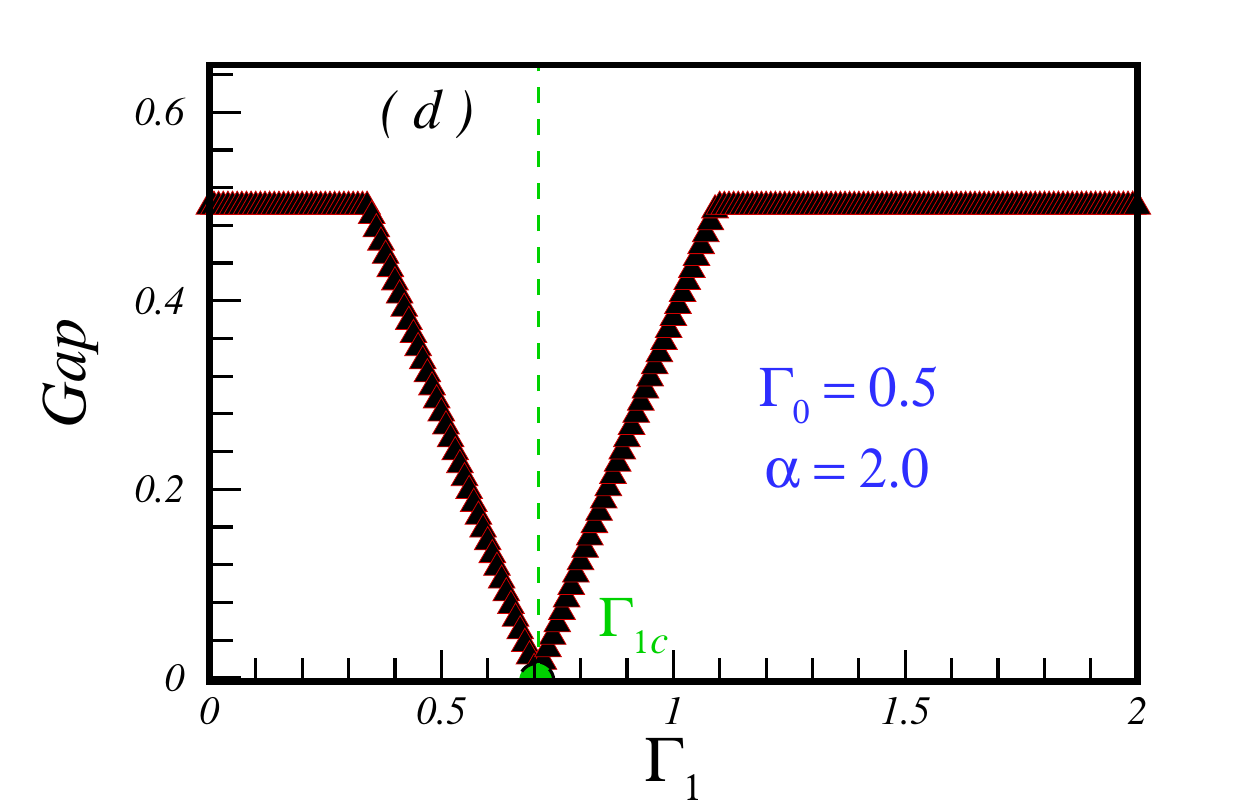}} 
	\centerline{\includegraphics[width=0.6\linewidth]{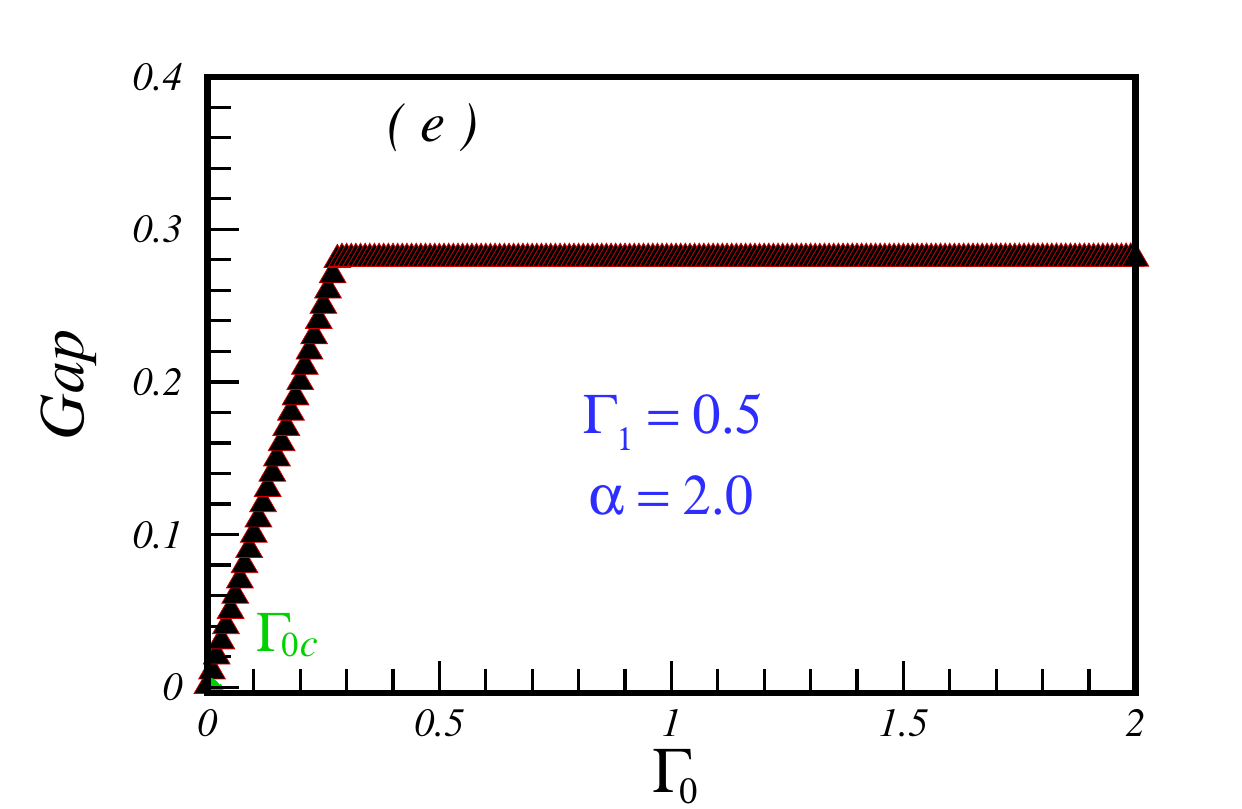}}
	\caption{The energy gap for a chain size $N=10000$. (a) In the case of  $\Gamma_1=0$, (b)  In the case of $\Gamma_0=0$, (c) When the system is first located in the gapless region of uniform GI, $\Gamma_0=0.5$ and $\alpha=-2.0$, (d) When the system is first located in the gapfull region of uniform GI, $\Gamma_0=0.5$ and $\alpha=2.0$, (e) When the system is first located in the gapless region of staggered GI, $\Gamma_1=0.5$ and $\alpha=2.0$. 
	}
	\label{Fig1-}
\end{figure}

We have calculated the energy gap for a chain size of $N=10000$ for different values of the Hamiltonian's parameters. Without losing generality, we have fixed $J=1$. The results are presented in Fig.~\ref{Fig1-}. The effect of the uniform GI, $\Gamma_1=0$, is shown in Fig.~\ref{Fig1-} (a). In complete agreement with previous works \cite{E5-5,E5-6}, the system remains gapless in the region  $\alpha\le \alpha_c=0$. As soon as the relative coefficient of off-diagonal exchange increases from zero, a gap opens and the system goes into a gapped phase. The opening of the gap in the vicinity of $\alpha_c=0$ scales with $\alpha^\nu$, where $\nu=1$.  The gap in the presence of a staggered GI is shown in Fig.~\ref{Fig1-} (b). As can be seen, the staggered GI does not induce a gap in the spectrum and the quantum system remains gapless.

We have studied the mutual effects of the uniform and staggered GIs on the energy gap of a spin-1/2 chain system. The results are shown in Fig.~\ref{Fig1-} (c), (d) and (e). In Fig.~\ref{Fig1-} (c), we see that when the system is in the gapless region, $\Gamma_0=0.5$ and $\alpha<\alpha_c=0$, the addition of the staggered GI  creates a gap in the spectrum that scales as $\Gamma_{1}^\nu$, where $\nu=1$. This indicates  that the system belongs to the  universality class of the transverse-field Ising model in one dimension. In Fig.~\ref{Fig1-} (d), we see that when the system is in the gapped region, $\Gamma_0=0.5$ and $\alpha>\alpha_c=0$, the presence of the staggered GI closes the gap at a critical point $\Gamma_{1c}(\alpha)$, which implies that there are two different gapped phases in this region. We observed a linear behavior of the energy gap in the critical region.   In Fig.~\ref{Fig1-} (e), we see that when the system is in the gapless region of the staggered GI, $\Gamma_1=0.5$, the introduction of the uniform GI induces a gap that behaves linearly near the critical point $\Gamma_{0c}=0$.

Order parameters are quantities that distinguish different regions of the ground state of a system by the degree of order they exhibit. An order parameter has a zero value in a disordered phase and a nonzero value in an ordered phase, and it reflects the symmetry (or the lack thereof) of the system. An order parameter is crucial for understanding quantum phase transitions, which are phase transitions that occur at zero temperature due to quantum fluctuations. An order parameter reveals the nature and the type of the quantum phase transition, which can be either continuous or discontinuous. In a continuous quantum phase transition, the order parameter changes smoothly, while in a discontinuous quantum phase transition, the order parameter changes abruptly.

\begin{figure}
	\centerline{\includegraphics[width=0.6\linewidth]{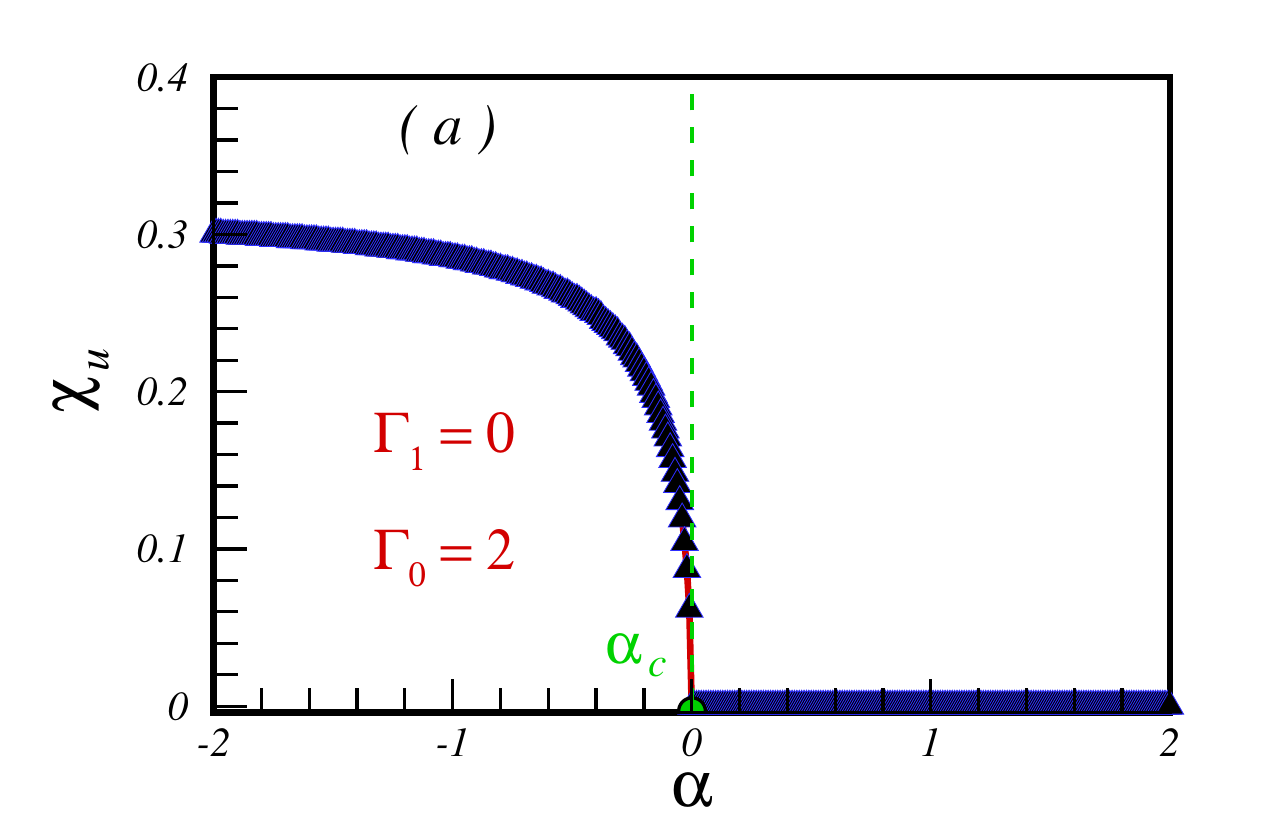}\includegraphics[width=0.6\linewidth]{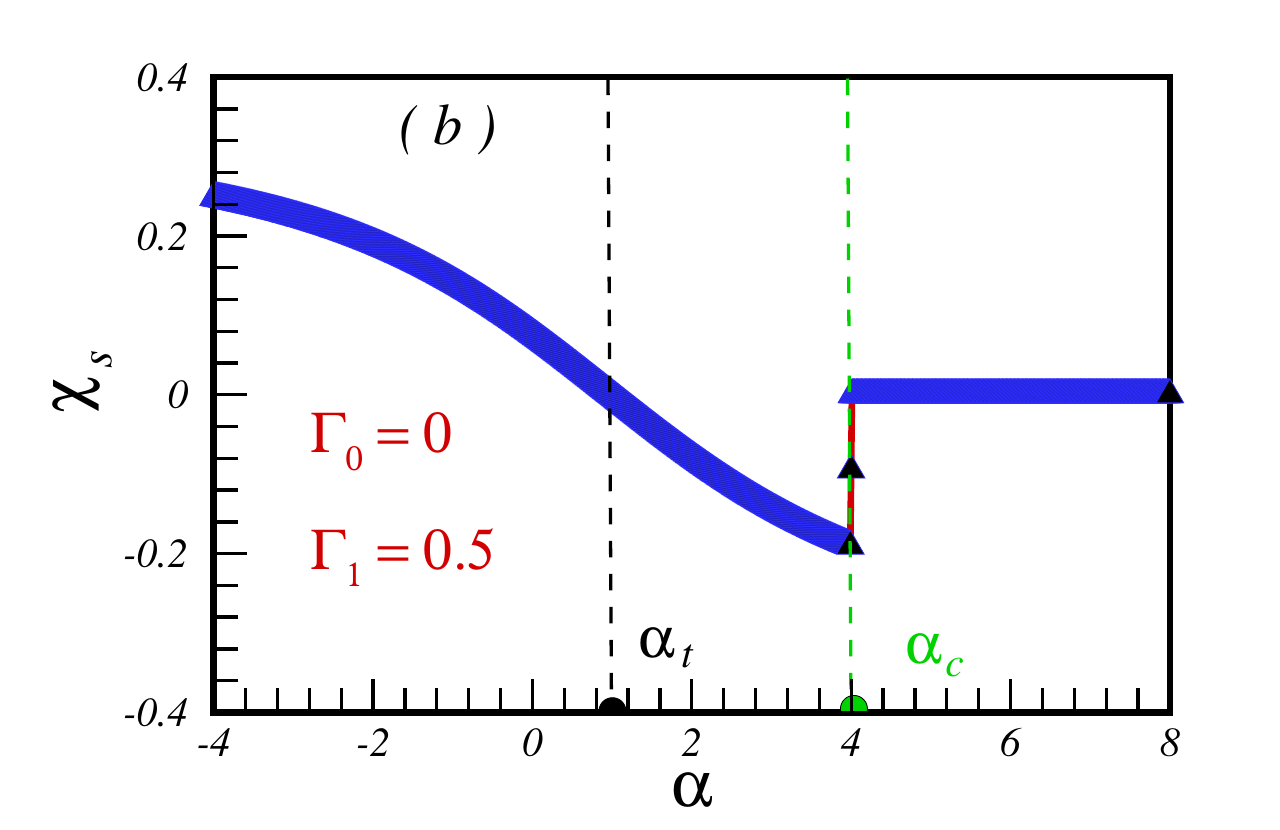}}
	\centerline{\includegraphics[width=0.6\linewidth]{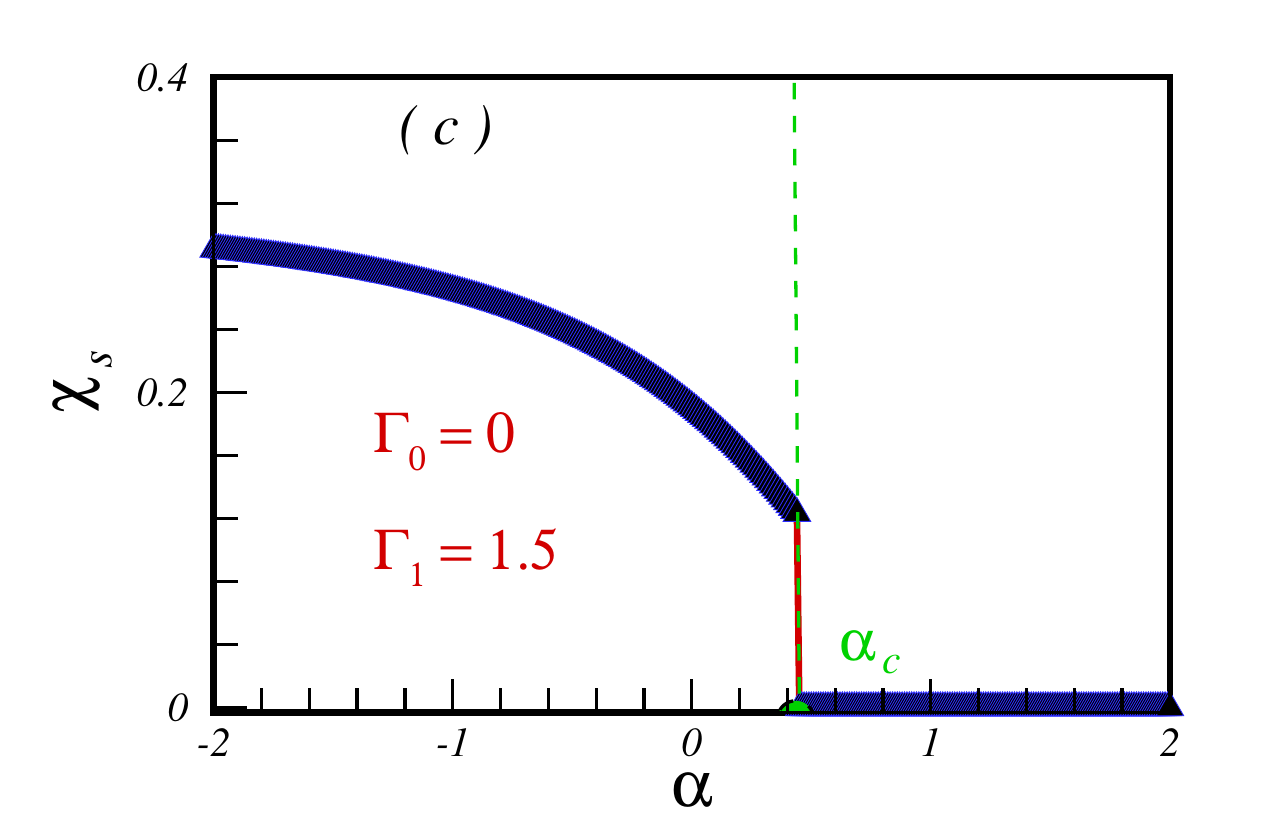}\includegraphics[width=0.6\linewidth]{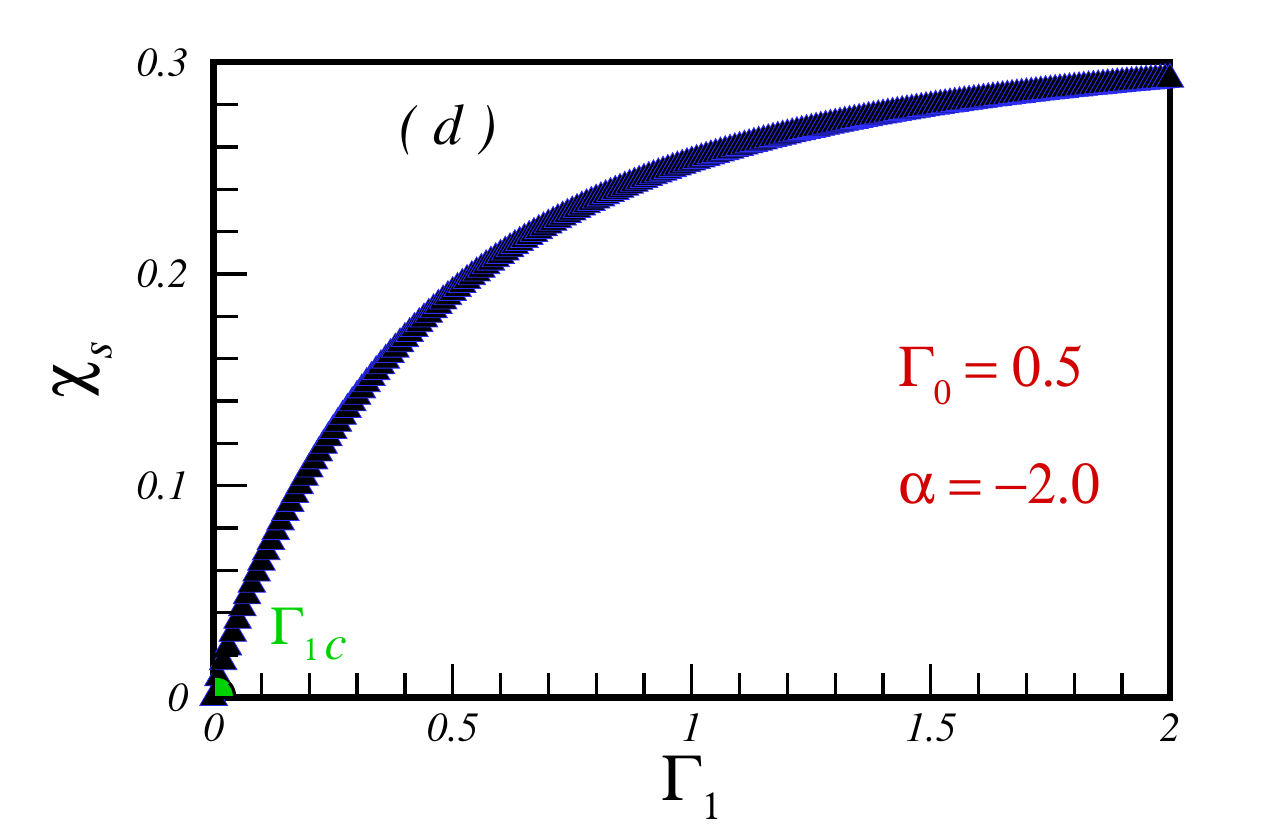}}
	\centerline{\includegraphics[width=0.6\linewidth]{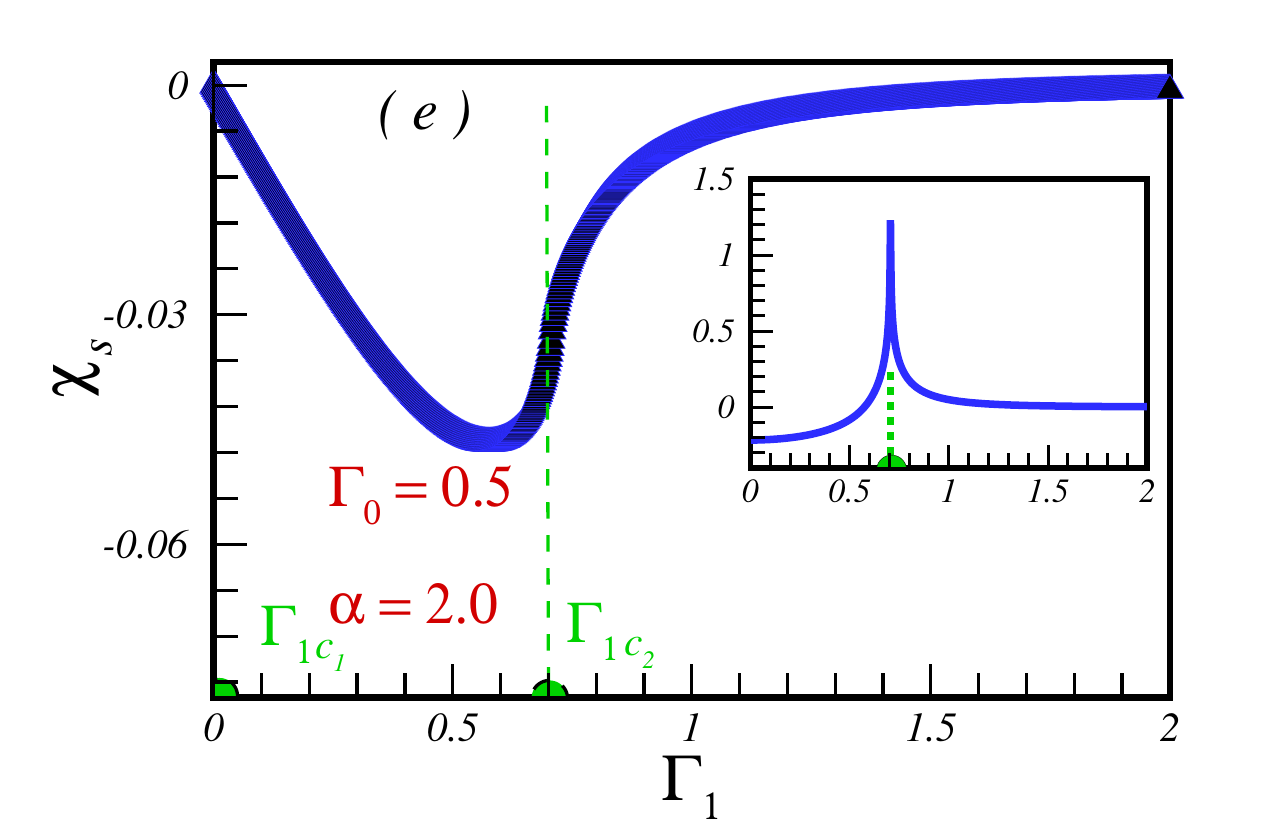}\includegraphics[width=0.6\linewidth]{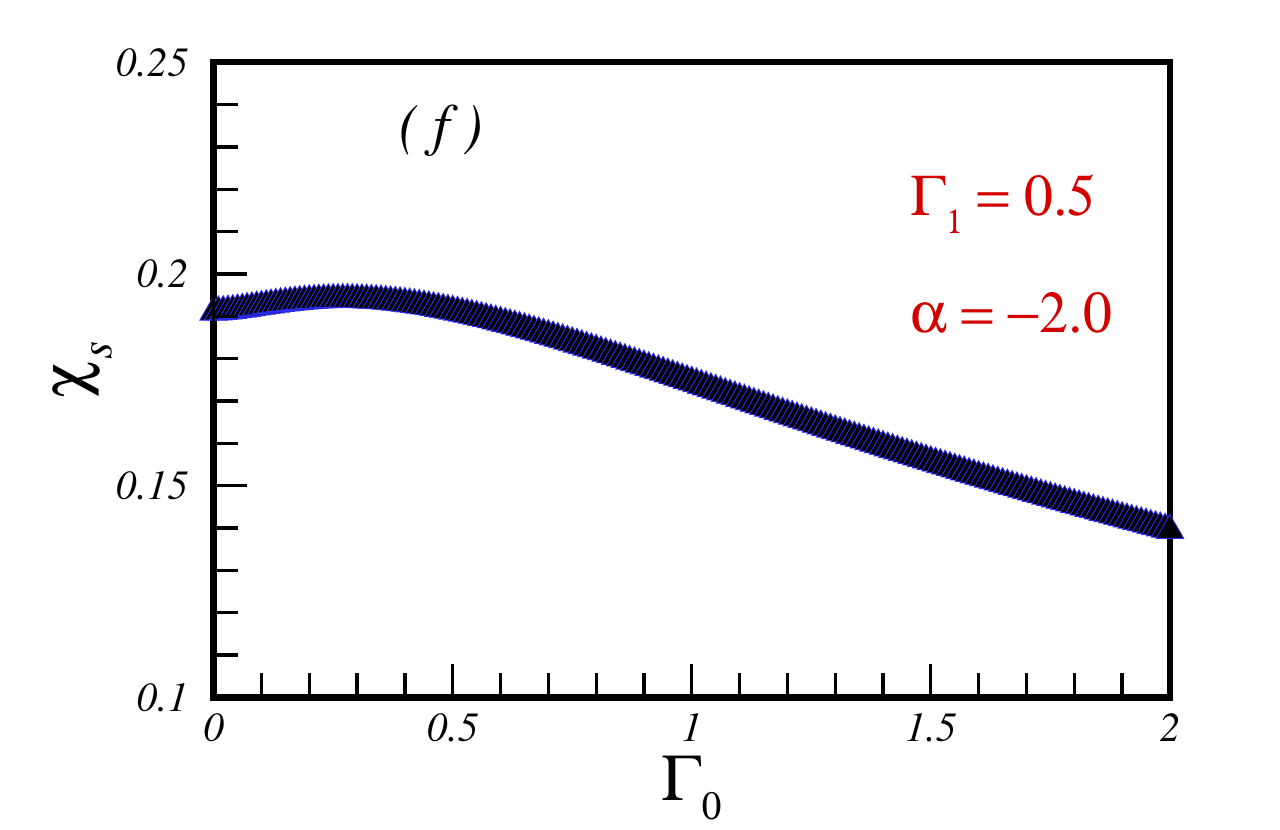}} 
	\caption{(a) The uniform chiral order parameter with respect to $\alpha$ in the absence of the staggered GI.  (b) The staggered chiral order parameter with respect to  $\alpha$ in the absence of the uniform GI in the region $\Gamma_1<J=1$. (c) The staggered chiral order parameter with respect to  $\alpha$ in the absence of the uniform GI and in the region $\Gamma_1>J=1$. (d) The staggered chiral order parameter with respect to $\Gamma_1$ in the presence of the uniform GI, $\Gamma_0=0.5$ and $\alpha=-2$. (e) The staggered chiral order parameter with respect to  $\Gamma_1$ in the presence of the uniform GI, $\Gamma_0=0.5$ and $\alpha=2$. In the inset, the first derivative of the staggered chiral order parameter with respect to the staggered  GI is plotted. (f) The staggered chiral order parameter with respect to  $\Gamma_0$ in the presence of the staggered GI, $\Gamma_1=0.5$ and $\alpha=-2$. 
	}
	\label{Fig2-}
\end{figure}

The chiral order parameter is a measure of the twist of the spins around a chain. 
Chiral ordering can lead to quantum phase transitions between different chiral phases or between chiral and non-chiral phases. A chiral phase is a state of matter that breaks time-reversal and reflection symmetries, but does not develop magnetic order even at zero temperature. 
The staggered chiral phase is recognized by staggered chiral order parameter defined as 
\begin{eqnarray}\label{eq8}
	\chi_s&=&\frac{1}{N} \sum_{n=1}^{N} (-1)^{n} \langle S _{n}^x S _{n+1}^y- S _{n}^y S _{n+1}^x  \rangle \nonumber \\
	&=& \frac{-i}{N} \sum\limits_{k>0} (1+\cos(k)) ~ \langle  a_k^\dag b_k+a_{-k}^\dag b_{-k} -H. c. \rangle \nonumber \\
	&+&\frac{1}{N} \sum\limits_{k>0} \sin(k)~ \langle  a_k^\dag b_k-a_{-k}^\dag b_{-k} +H. c. \rangle,
\end{eqnarray}
where $\langle ... \rangle$ denotes the expectation value on the ground state of $H_k$. In fact $\chi$ is a measure of the non-coplanarity of neighboring spins. It should be noted that by removing the $(-1)^n$, $\chi_u$ is known as the uniform chiral order parameter.  

We have computed the staggered and uniform chiral order parameter in our model, which are shown in Fig.~\ref{Fig2-}. The GI can induce chiral ordering in the ground state phase diagram of spin-1/2 chains, depending on its strength. When the staggered GI is absent, $\Gamma_1=0$, the ground state exhibits long range chiral order in the region $\alpha<\alpha_c=0$, as seen in Fig.~\ref{Fig2-} (a). When $\alpha>0$, the chiral order vanishes, indicating that the uniform chiral order parameter characterizes the long range order in the region $\alpha<0$.  The emergence of a chiral ordering near the critical point, where $ \alpha_c = 0 $, follows the scaling relation $\chi_u \sim |\alpha|^\beta $ with the critical exponent $ \beta = 1/2 $. 

As shown in Fig.~\ref{Fig2-} (b) and (c), the staggered GI has a completely different effect. We identified two distinct regions depending on the value of $\Gamma_1$. For $\Gamma_1<J=1$, the system exhibits long-range staggered chiral order in the region $\alpha<0$, as seen in Fig.~\ref{Fig2-} (b).  At $\alpha = 1$, the Gamma interaction becomes isotropic, leading to $\langle S_n^x S_{n + 1}^y \rangle = \langle S_n^y S_{n + 1}^x \rangle$. Consequently, the staggered chiral order parameter becomes zero. In fact, by passing the line $\alpha=1$, the sign of the staggered chiral order parameter switches from positive to negative. As $\alpha$ increases further, a first order quantum phase transition occurs at a specific value of the relative coefficient of off-diagonal exchange, $\alpha_{c}(\Gamma_1)$. For $\Gamma_1>J=1$, the system is in the staggered chiral phase in the region $\alpha<\alpha_{c}(\Gamma_1)$, as shown in Fig.~\ref{Fig2-} (c). A first order phase transition takes place at this critical point.

The paragraph describes the effects of uniform and staggered GIs on a system in different phases, as shown in Fig.~\ref{Fig2-} (d), (e) and (f). The system is in the uniform chiral phase when $\Gamma_0=0.5$ and $\alpha=-2.0$, as seen in Fig.~\ref{Fig2-} (d). If the staggered Gamma term is added, the system transitions to the staggered chiral phase immediately as a sign  of the second order phase transition. However, when the system is not in the uniform chiral phase, $\Gamma_0=0.5$ and $\alpha=2.0$,  as shown in Fig.~\ref{Fig2-} (e), as soon as the staggered Gamma applies, the system goes to the negative staggered chiral phase and by more increasing the $\Gamma_1$, system undergoes a second order quantum phase transition at $\Gamma_{1c_{2}} (\alpha)$ (see also inset). Lastly, when the system has a staggered GI, $\Gamma_1=0.5$ and $\alpha=-2.0$, as in Fig.~\ref{Fig2-} (f), adding the uniform GI does not cause any quantum phase transition, and the system stays in the staggered chiral phase.

Here we explain the concept of quantum spin nematic phase, which is a state of matter where the electron spins in a chain point along two different directions, instead of one. This breaks the rotational symmetry of the system, but not the time-reversal symmetry. This phase can occur in spin-1/2 chains with some kinds of interactions, such as the biquadratic XY model with rhombic single-ion anisotropy \cite{R1-6}. GI can also cause nematic phase in spin-1/2 chains in some situations, such as near the antiferromagnetic Kitaev region with nonzero GI \cite{R1-7}.
The spin nematic order parameter is a quantity that measures the degree of spin quadrupolar order in a system. Spin quadrupolar order means that the spins point along two orthogonal axes, rather than one. The spin nematic order parameter is usually a second-rank, traceless, symmetric tensor, such as $Q_{\alpha\beta} = \langle S_\alpha S_\beta + S_\beta S_\alpha \rangle$ where $S_\alpha$ ($\alpha = x, y, z$) are the spin operator components.  Here, we computed the staggered nematic order parameter defined as
\begin{eqnarray}\label{eq8}
	Q_s&=&\frac{1}{N} \sum_{n=1}^{N} (-1)^{n} \langle S _{n}^x S _{n+1}^y+ S _{n}^y S _{n+1}^x  \rangle \nonumber \\
	&=& \frac{-i}{N} \sum\limits_{k>0} (1+\cos(k)) ~ \langle  a_k^\dag b_{-k}^\dag+a_{-k}^\dag b_{k}^\dag -H. c. \rangle \nonumber \\
	&-&\frac{1}{N} \sum\limits_{k>0} \sin(k)~ \langle  a_k^\dag b_{-k}^\dag-a_{-k}^\dag b_{k}^\dag +H. c.  \rangle,
\end{eqnarray}
please note that by removing the $(-1)^n$, $Q_u$ is known as the uniform nematic order parameter. It is imperative to note that while the traditional single-site nematic order parameter is not applicable to spin-1/2 systems due to the absence of single-ion anisotropy, the concept of bond nematicity is relevant. Our definition of the staggered nematic order parameter is based on intersite spin correlations, reflecting the collective behavior of spin pairs rather than individual spin sites. 

\begin{figure}
	\centerline{\includegraphics[width=0.6\linewidth]{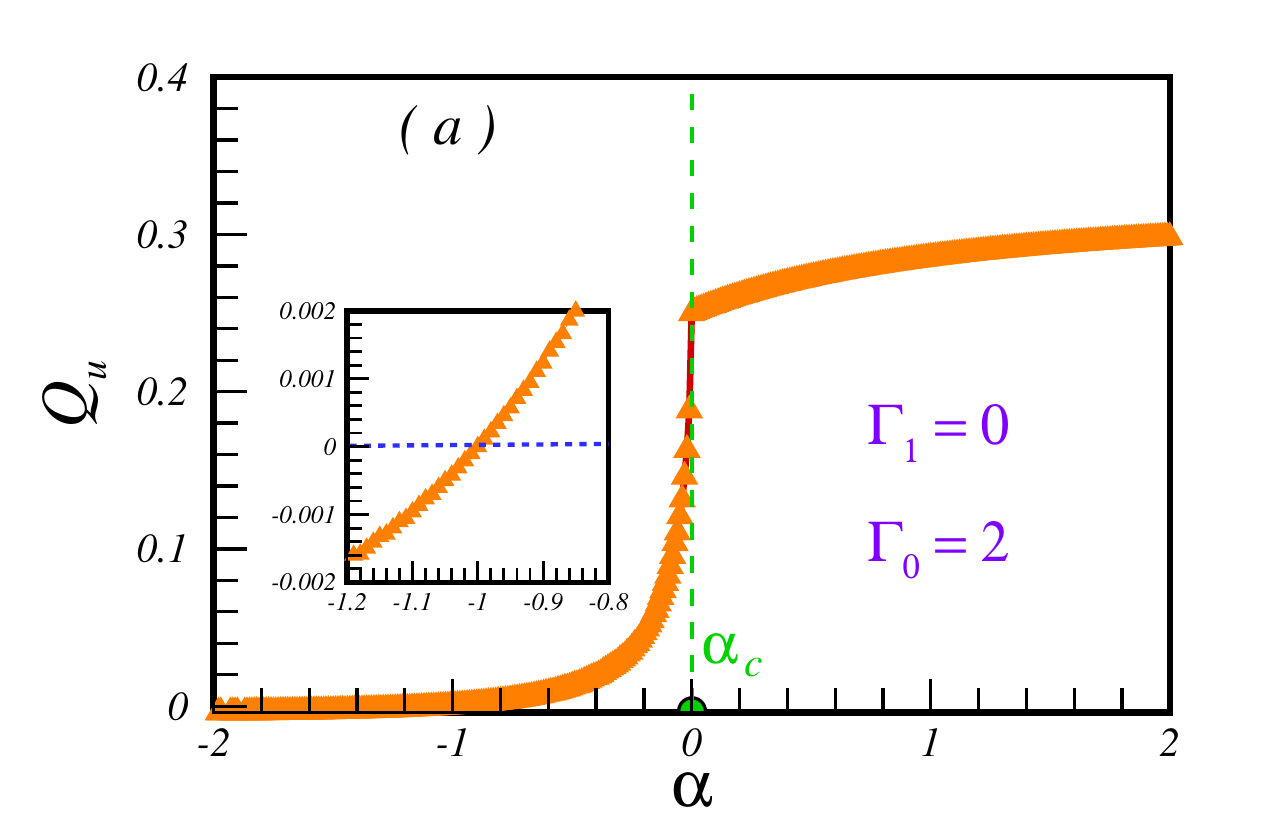}\includegraphics[width=0.6\linewidth]{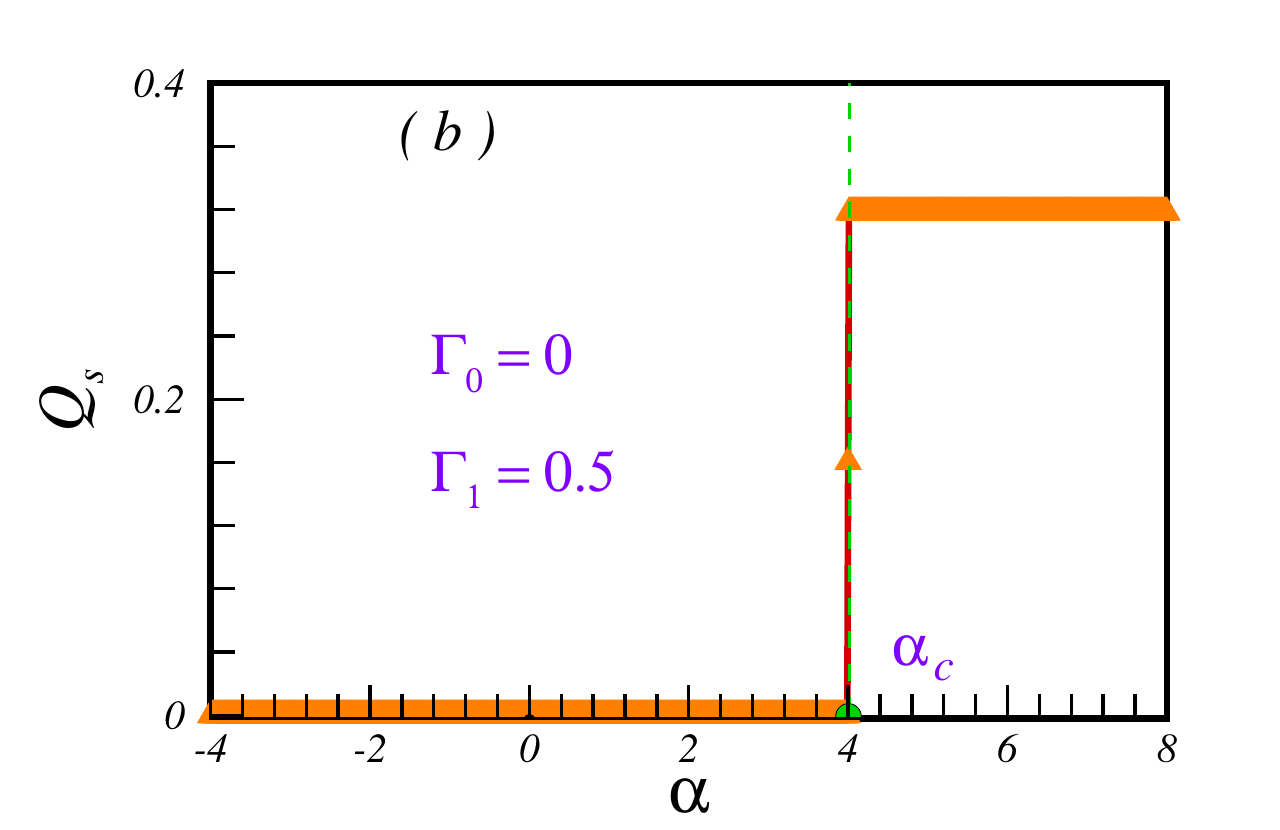}}
	\centerline{\includegraphics[width=0.6\linewidth]{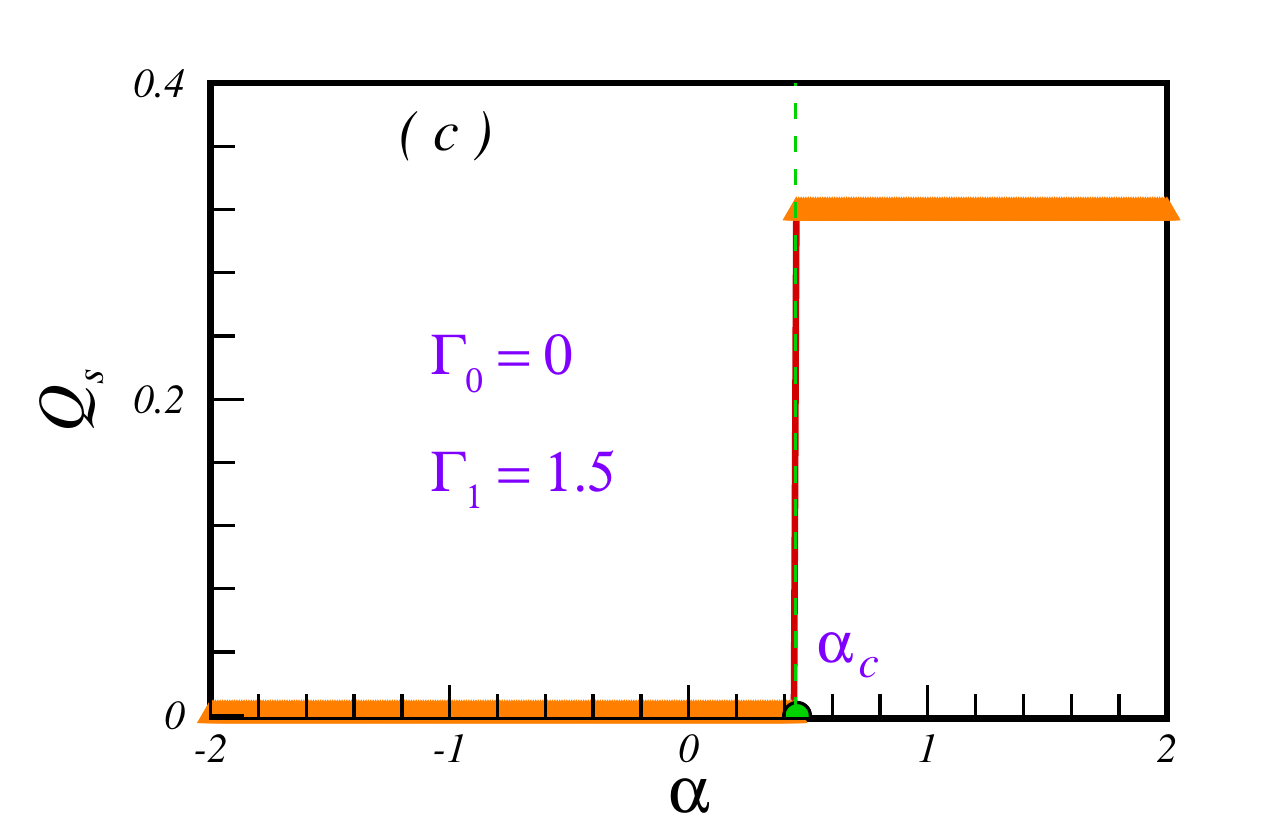}\includegraphics[width=0.6\linewidth]{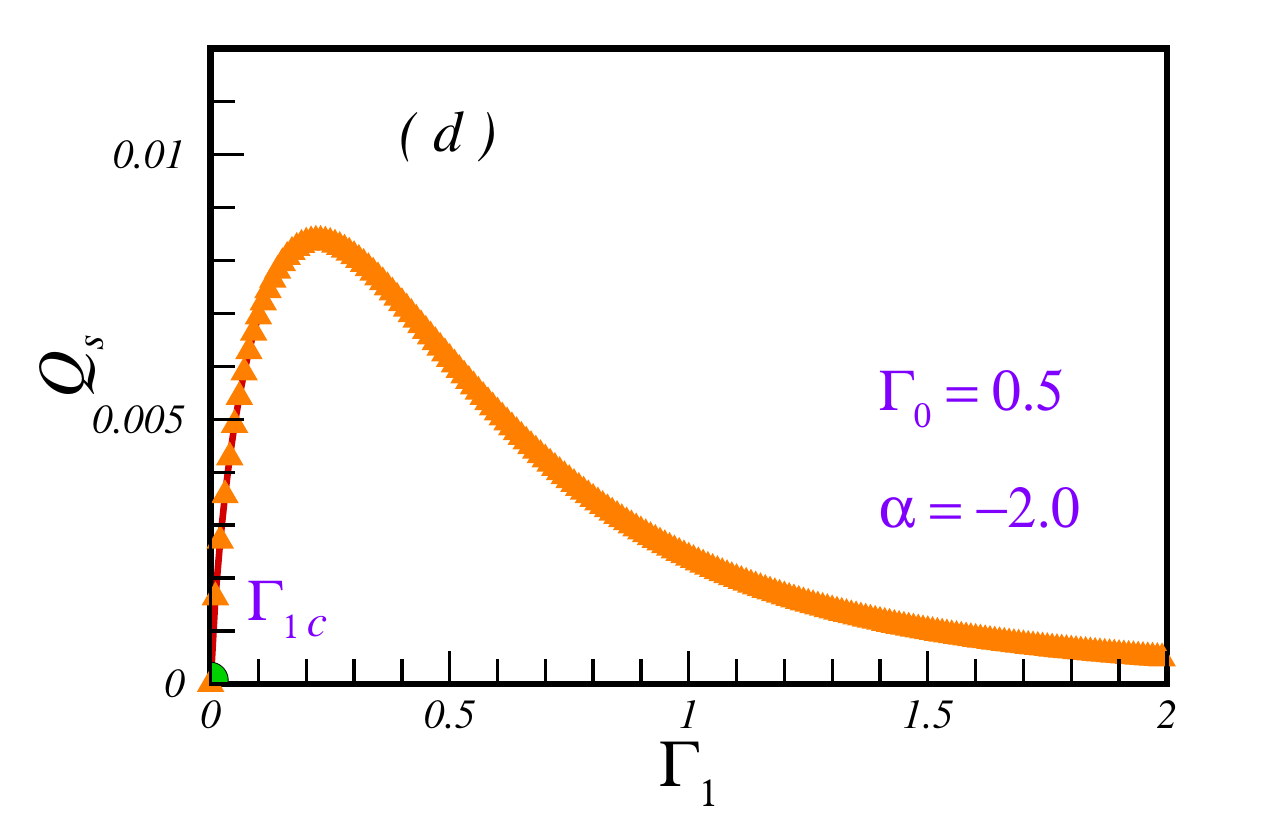}}
	\centerline{\includegraphics[width=0.6\linewidth]{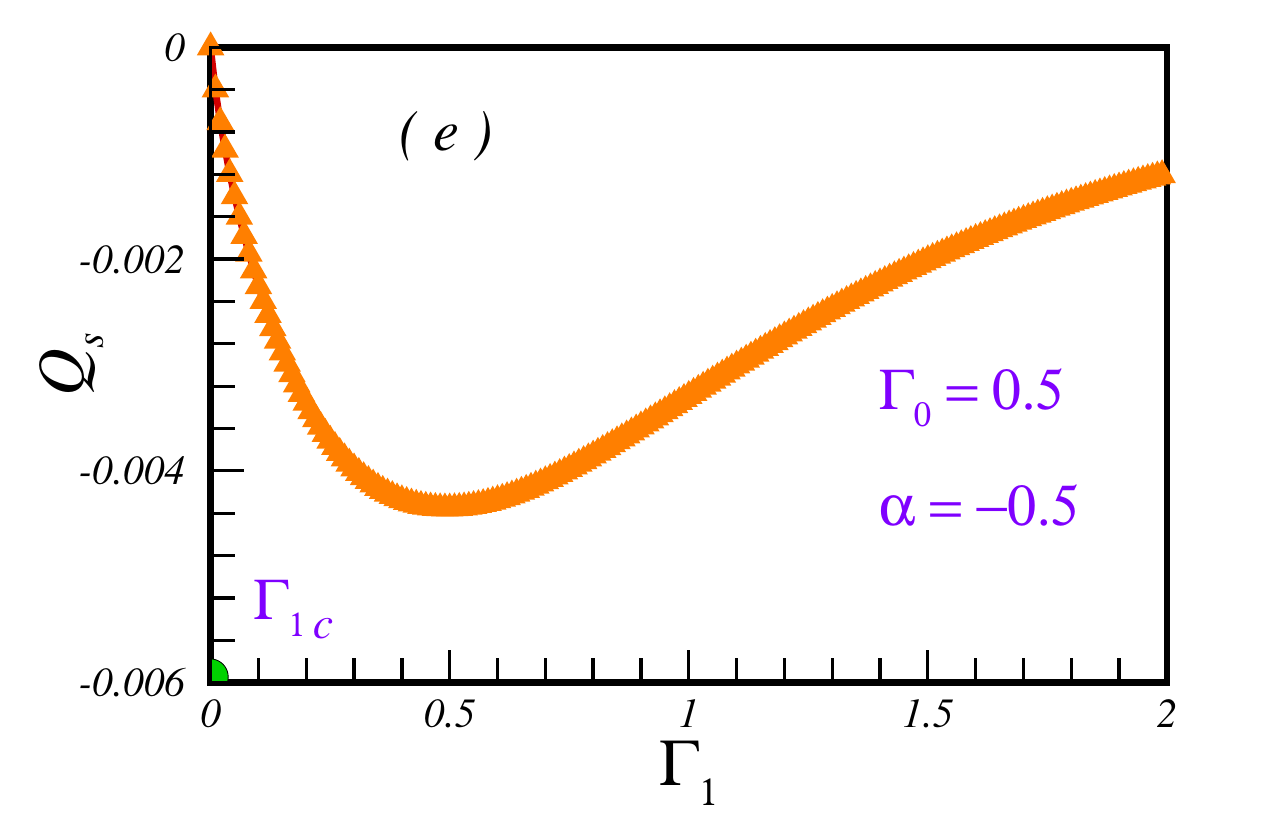}\includegraphics[width=0.6\linewidth]{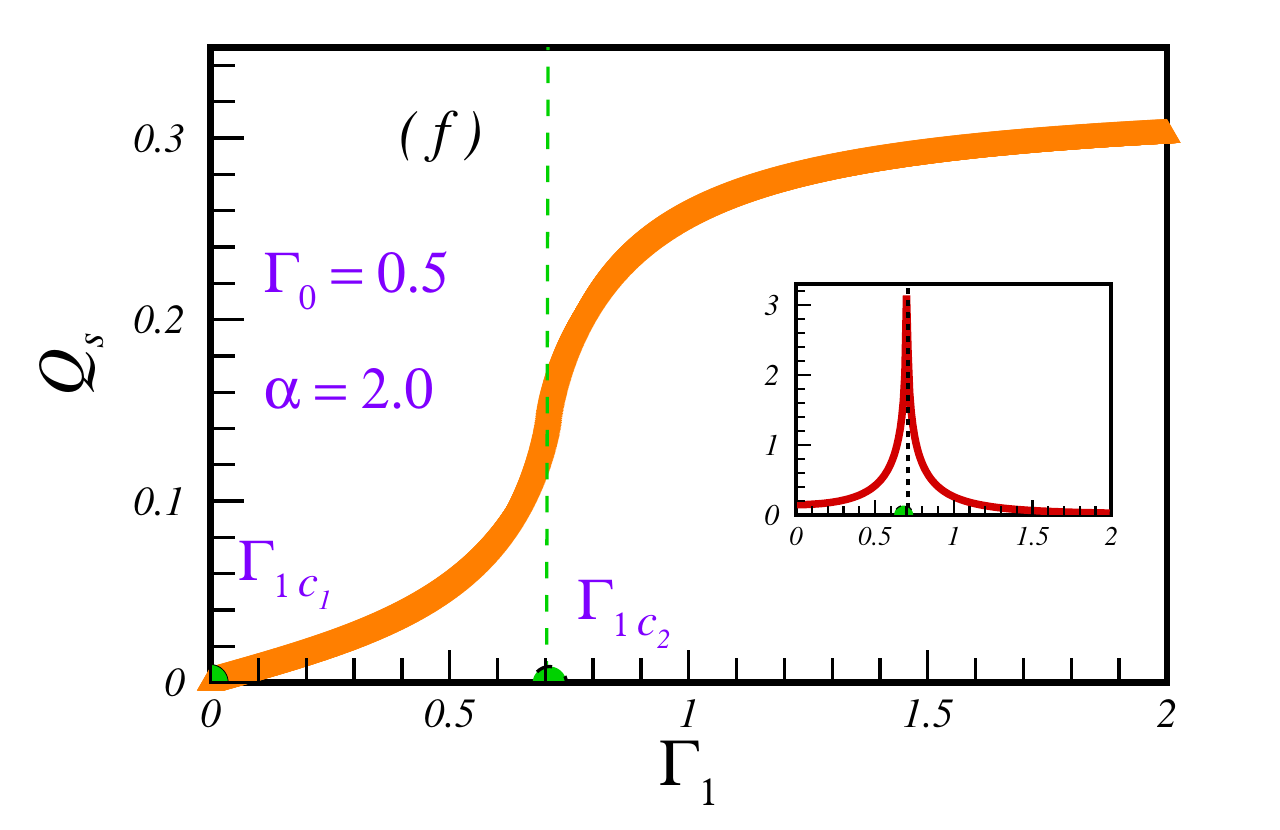}} 
	\caption{(a) The uniform nematic order parameter with respect to  $\alpha$ in the absence of the staggered GI.  (b) The staggered nematic order parameter with respect to  $\alpha$ in the absence of the uniform GI and in the region $\Gamma_1<J=1$. (c) The staggered nematic order parameter with respect to  $\alpha$ in the absence of the uniform GI and in the region $\Gamma_1>J=1$. (d) The staggered nematic order parameter with respect to  $\Gamma_1$ in the presence of the uniform GI, $\Gamma_0=0.5$ and $\alpha=-2$. (e) The staggered nematic order parameter with respect to  $\Gamma_1$ in the presence of the uniform GI, $\Gamma_0=0.5$ and $\alpha=-0.5$. (f) The staggered nematic order parameter with respect to  $\Gamma_1$ in the presence of the uniform GI, $\Gamma_0=0.5$ and $\alpha=2$. In the inset, the first derivative of the staggered nematic order parameter with respect to the staggered  GI is plotted.   
	}
	\label{Fig3-}
\end{figure}

The paragraph reports the results on the nematic order parameter, which is shown in Fig.~{\color{blue}\ref{Fig3-}}. As is seen in Fig.~\ref{Fig3-} (a), the system is in the uniform nematic phase when $\alpha>0$. 
In the region $\alpha<0$, the uniform nematic order parameters remains non zero and positive until $\alpha=-1$. As soon as the $\alpha$ decreases from $-1$, the uniform nematic order parameters becomes negative and its domain increases as is seen clearly in the inset of Fig.~\ref{Fig3-} (a).  If the staggered Gamma term is added, $\Gamma_0=0$ and $\Gamma_1>0$, the system transitions to the staggered nematic phase at $\alpha_{c} (\Gamma_{1})$, as seen in Fig.~\ref{Fig3-} (b) and (c). This is a first order quantum phase transition. The effect of the staggered GI on the system with a uniform GI is studied. As in Fig.~\ref{Fig3-} (d) is seen, a quantum phase transition into the staggered spin nematic  is seen  when the uniform  Gamma term is added and $\alpha=-2.0<-1$. On the other hand, as in Fig.~\ref{Fig3-} (e) is observed, a quantum phase transition into the negative staggered spin nematic  happens  when the uniform  Gamma term is added and  $-1<\alpha=-0.5<0$. Lastly, Fig.~\ref{Fig3-} (f) shows that when the system has a uniform GI, $\Gamma_0=0.5$, and $\alpha=2$, the system enters the staggered nematic phase at $\Gamma_{1c_{1}}=0$ as soon as the staggered Gamma term is applied. As the staggered Gamma increases, the system undergoes a second order quantum phase transition at $\Gamma_{1c_{2}} (\alpha)$. The inset shows the first derivative of the staggered nematic order parameter versus $\Gamma_1$. A discontinuity at the critical point $\Gamma_{1c_{2}} (\alpha)$ indicates the second order phase transition.

It is known that the dimer order parameter can be used to characterize the quantum phase transitions and the nature of the phases in spin systems with competing interactions. 
Mathematically, it can be written as

\begin{eqnarray}\label{eq8}
	D&=&\frac{1}{N} \sum_{n=1}^{N} (-1)^{n} \langle S _{n}^x S _{n+1}^x+ S _{n}^y S _{n+1}^y  \rangle \nonumber \\
	&=& \frac{1}{N} \sum\limits_{k>0} (-1+\cos(k)) ~ \langle  a_k^\dag b_k+a_{-k}^\dag b_{-k} +H. c. \rangle \nonumber \\
	&+&\frac{i}{N} \sum\limits_{k>0} \sin(k)~ \langle  a_k^\dag b_k-a_{-k}^\dag b_{-k} -H. c. \rangle.
\end{eqnarray}
A nonzero value of $D$ indicates the presence of a dimer phase, which is a gapped quantum phase of matter with no magnetic order.

\begin{figure}
	\centerline{\includegraphics[width=0.55\linewidth]{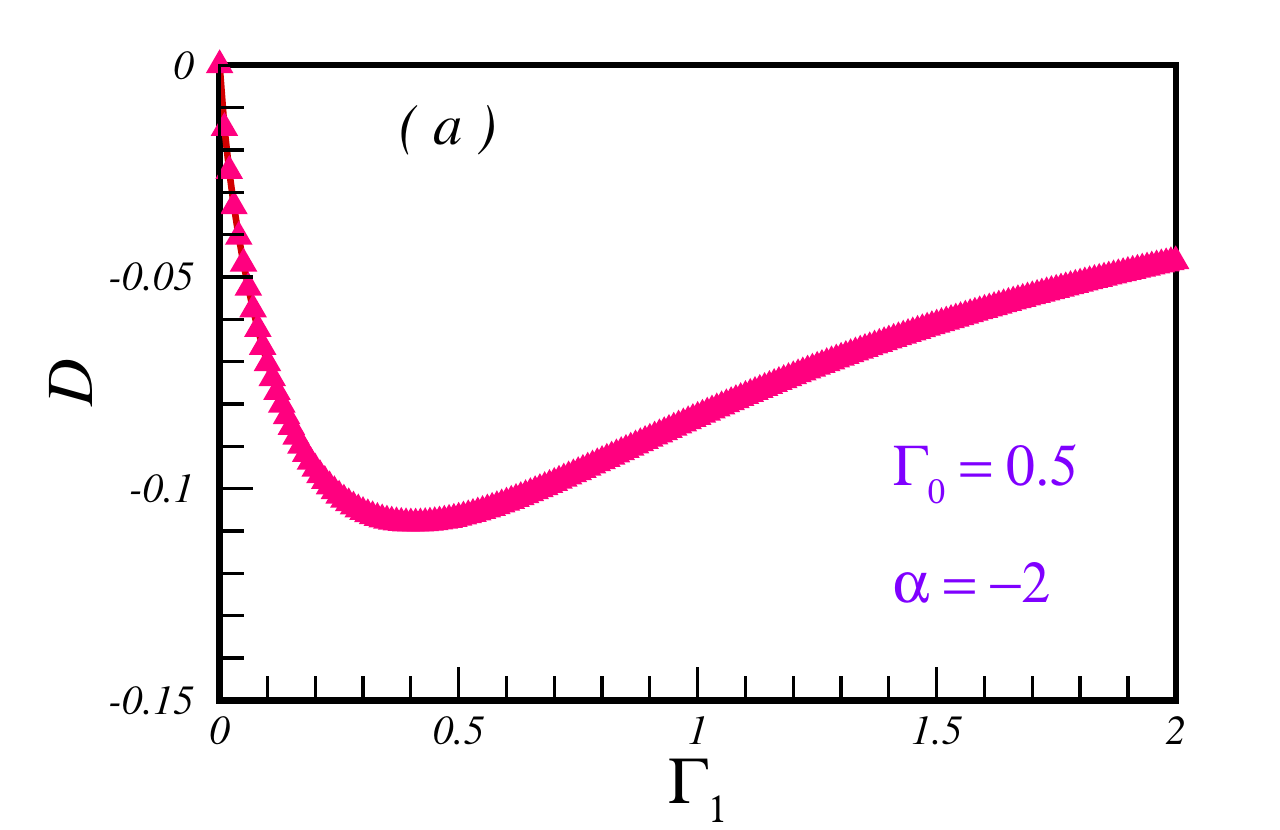}\includegraphics[width=0.55\linewidth]{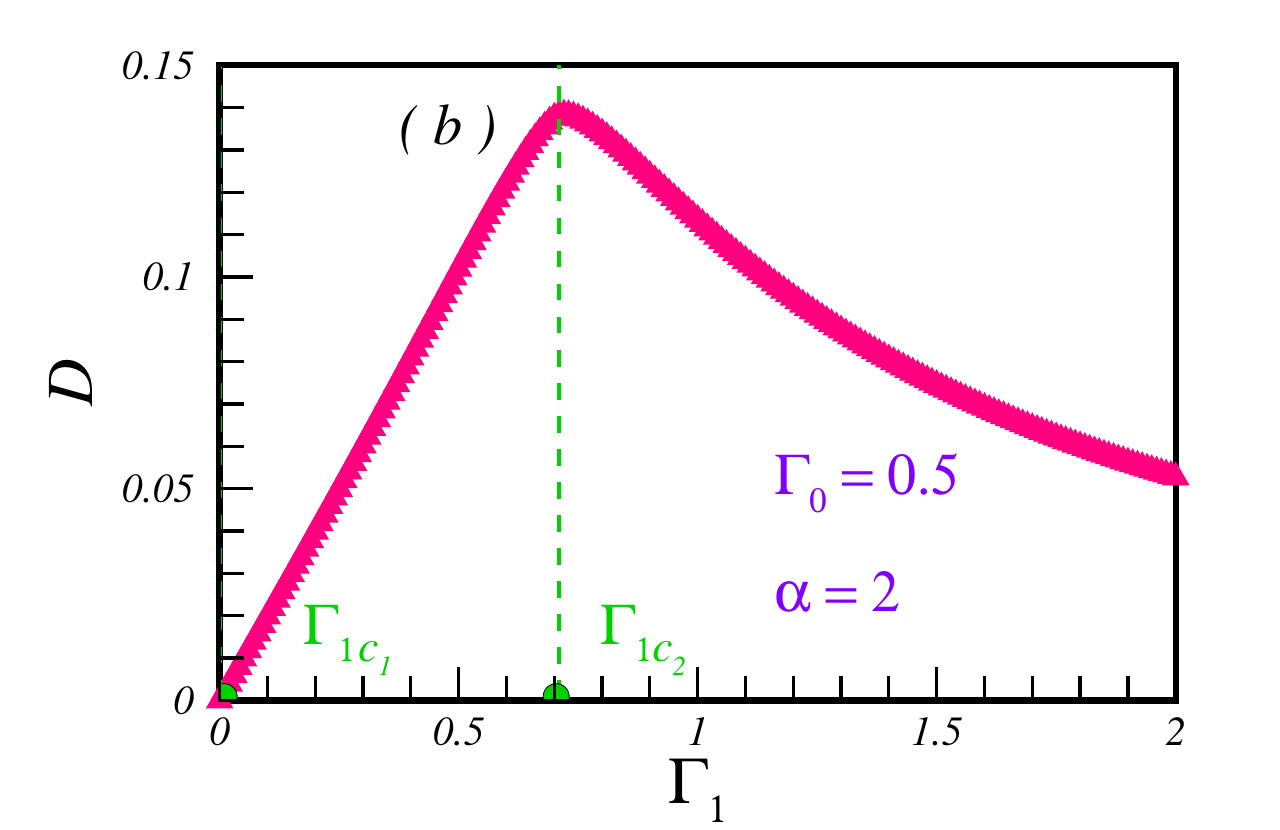}}
	\caption{(a) The dimer order parameter with respect to  $\Gamma_1$ in the presence of the uniform GI, $\Gamma_0=0.5$ and $\alpha=-2$. (b) The dimer order parameter with respect to  $\Gamma_1$ in the presence of the uniform GI, $\Gamma_0=0.5$ and $\alpha=2$. Positive results show odd-parity dimer ordering. 
	}
	\label{Fig4-}
\end{figure}

Here, we have calculated the dimer order parameter for different values of the GIs and results are presented in Fig.~\ref{Fig4-}. First of all, we did not find any dimerization in the presence of the pure uniform or staggered GI as expected. However, for modulated GIs, we find different dimer phases depending on the sign of $\alpha$. In Fig.~\ref{Fig4-} (a), we see that the system has long-range dimer ordering for $\alpha<0$. Fig.~\ref{Fig4-} (b), we see that the system has odd-parity dimer ordering for $\alpha>0$. This means that the ground states are dimer states with the unit $\frac{1}{\sqrt{2}} (\ket{\uparrow \downarrow} +  \ket{ \downarrow \uparrow} )$, which has even parity under bond inversion, unlike the singlet dimers with odd parity. We note that the odd-parity dimer ordering was also reported in some frustrated spin-1/2 XXZ chains with competing ferromagnetic and antiferromagnetic interactions \cite{R1-8}.

\section{Conclusion}

Our study explored the one-dimensional spin-1/2 XX model with a modulated Gamma model. The spin-1/2 XX chain model is a basic but significant model in quantum physics that describes a one-dimensional system of interacting spin-1/2 particles with nearest-neighbor exchange interactions.

The GI in spin-1/2 chains is caused by the bond-dependent magnetic couplings that result from the spin-orbit coupling and the crystal field effects in some materials. The spin-orbit coupling is a relativistic effect that links the spin and orbital angular momentum of the electrons, while the crystal field effects are the deformations of the electron orbitals due to the surrounding ions. These effects can lead to anisotropic and direction-dependent exchange interactions between the spins, such as the Gamma terms. The GI is an off-diagonal term that flips the spins along the x and y directions. The GI can also be seen as a special case of the Dzyaloshinskii-Moriya interaction, which is a bond-dependent antisymmetric exchange interaction that arises from the spin-orbit coupling and the absence of inversion symmetry.

We defined the modulated Gamma as a superposition of uniform and 
staggered GIs and used the fermionization technique for exactly diagonalizing the Hamiltonian. 

First, we focused on the energy gap. We found that, while the pure staggered GI cannot induce a gap in the spectrum, the pure uniform GI induces a gap in the region where the relative coefficient of off-diagonal exchange, $\alpha$, is positive. In this gapped region, if we add the staggered GI, it closes the gap at a certain critical value of strength of staggered GI.

Second, we investigated several functions such as, uniform and staggered chiral order parameter, uniform and staggered nematic order parameter and dimer order parameter. In the presence of the pure uniform GI, the system is in the gapped spin nematic phase in the region $\alpha>0$. In the case where $\alpha<0$ we found two gapless composite  phases separated by a  line at $\alpha=-1$. In the region $-1<\alpha<0$, system is in a composite phase consists of chiral and weak spin nematic ordering. On the other hand, in the region  $\alpha<-1$, system is in a composite phase with  chiral and weak  negative spin nematic ordering. The results on the pure staggered Gamma were completely different. In the region $\Gamma_1>J=1$, the system is in the gapless staggered chiral phase in the region $\alpha<0$ and a second order phase transition occurs into the gapless staggered nematic phase at $\alpha_c=0$. In the region where $\Gamma_1<J=1$, the system is in the gapless staggered chiral phase in the region $\alpha<1$ and goes to another gapless phase with negative staggered chiral ordering at $\alpha=1$. The system remains in this phase up to $\alpha_{c}(\Gamma_1)$, where a first order quantum phase transition occurs into the gapless staggered nematic phase.

Third, we have demonstrated that when the GI has both uniform and staggered components, the ground state phase diagram has four composite phases:\\
\\
$\ast$ For $-1<\alpha<0$, the system is in a gapped composite phase with dimer, negative staggered nematic and staggered chiral order.
\\
\\
$\ast$ For $\alpha<-1$, the system is in a gapped composite phase with dimer, weak staggered nematic and staggered chiral order.
\\
\\
$\ast$ For $0<\alpha<1$ and $\Gamma_1<\Gamma_{1c} (J,\Gamma_0)$, the system is in a gapped composite phase with odd-parity dimer and staggered chiral order and staggered spin nematic order. In the region $0<\alpha<1$ and $\Gamma_1>\Gamma_{1c}(J,\Gamma_0)$, the system is in a gapped composite phase with odd-parity dimer and staggered spin nematic order.\\
\\
$\ast$ For $1<\alpha<\alpha_c$ and $\Gamma_1<\Gamma_{1c}(J,\Gamma_0)$, the system is in a gapped composite phase with odd-parity dimer and negative staggered chiral order and staggered spin nematic order. In the region $1<\alpha<\alpha_c$ and $\Gamma_1>\Gamma_{1c}(J,\Gamma_0)$, the system is in a gapped composite phase with odd-parity dimer and staggered spin nematic order. \\
\\
$\ast$ For $\alpha>\alpha_c$, the system is in a gapped composite phase with odd-parity dimer and staggered nematic order. \\

Our study of the spin-1/2 XX chain with a modulated Gamma interaction contributes to the fundamental understanding of quantum phase transitions, particularly in systems that exhibit various  characteristics. The exact diagonalization of the Hamiltonian using the Fermionization technique not only enriches the theoretical framework but also provides a solid foundation for exploring quantum criticality in low-dimensional systems.

The identification of gapped and gapless regions, along with the examination of various order parameters, sheds light on the intricate ground state phase diagram of the model. This is particularly relevant for the development of quantum materials, where understanding the nature of phase transitions can lead to the discovery of new states of matter with potential applications in quantum computing and information storage.

Moreover, the observation of first, second order, gapless-gapless, and gapped-gapped phase transitions opens avenues for experimental verification. Such transitions are pivotal for the realization of robust quantum states that are protected against local perturbations, making them ideal candidates for quantum error correction schemes.

In conclusion, our results not only advance the theoretical landscape of quantum phase transitions but also have practical implications for the design and manipulation of quantum systems. We believe that our work will inspire further theoretical and experimental studies, potentially leading to breakthroughs in quantum technologies.

\nolinenumbers


\clearpage

\begin{thebibliography}{99}
	
	\bibitem{E1} A.N. Vasiliev, O.S. Volkova, E.A. Zvereva, M.M. Markina, {\it Low-dimensional magnetism}, CRC Press, ISBN 9781032239002 (2019).
	
	\bibitem{E2} A. Vasiliev, O. Volkova, E. Zvereva and M. Markina, {\it Milestones of low-D quantum magnetism}, npj Quant. Mater. {\bf 3}, 18 (2018), \doi{10.1038/s41535-018-0090-7}.
	
	
	
	\bibitem{E3-0} U. Schollwöck , J. Richter , D.J.J. Farnell , R.F. Bishop (Eds.), {\it Quantum Magnetism}, Lecture Notes in Physics, (Springer, Berlin, 2004), \doi{10.1007/b96825}.
	
	
	\bibitem{E3} R. J. Baxter, {\it Exactly Solved Models in Statistical Mechanics}, Academic Press, 1982 Copyright: Rodney J. Baxter (2004), \doi{10.1142/9789814415255_0002}.
	
	
	\bibitem{E4-0}
	A. Gelfert and W. Nolting,  {\it The absence of finite-temperature phase transitions in low-dimensional many-body models: a survey and new results}, J. Phys.: Condens. Matter {\bf 13}, R505 (2001), \doi{10.1088/0953-8984/13/27/201}.
	
	
	
	
	\bibitem{E4}
	M. Takahashi,  {\it Thermodynamics of one-dimensional solvable models},Cambridge: Cambridge University Press, (1999).
	
	\bibitem{E5}
	S. Sachdev, {\it Quantum Phase Transitions}, 2nd ed. Cambridge University Press, (2011).
	
	
	\bibitem{E5-0}
	E. Lieb, T. Schultz  and D. Mattis, {\it Two soluble models of an antiferromagnetic chain}, Ann. Phys. {\bf 16}, 407 (1961), \doi{10.1016/0003-4916(61)90115-4}.
	\bibitem{E5-0-p1}
	 S. Katsura, {\it Statistical Mechanics of the Anisotropic Linear Heisenberg Model}, Phys. Rev. {\bf 127}, 1508 (1962); 129, 2835 (1963), \doi{10.1103/PhysRev.127.1508}.
	
	\bibitem{E5-1} E. Barouch, B. M. McCoy and M. Dresden, {\it Statistical Mechanics of the XY Model}. I, Phys. Rev. A {\bf 2}, 1075 (1970), \doi{10.1103/PhysRevA.2.1075}.
	\bibitem{E5-1-p1}
	 E. Barouch and B. M. McCoy, {\it Statistical Mechanics of the XY Model. II. Spin-Correlation Functions}, Phys. Rev. A {\bf 3}, 786 (1971), \doi{10.1103/PhysRevA.3.786}.
	
	\bibitem{E5-2} A. Kitaev, {\it Anyons in an exactly solved model and beyond}, Ann. Phys. {\bf 321}, 2 (2006), \doi{10.1016/j.aop.2005.10.005}.
	
	\bibitem{E5-3} L. Janssen, E. C. Andrade and M. Vojta, {\it Magnetization processes of zigzag states on the honeycomb lattice: Identifying spin models for $\alpha-RuCl_3$ and Na$_2$IrO$_3$}, Phys. Rev. B {\bf 96}, 064430 (2017), \doi{10.1103/PhysRevB.96.064430}.
	
	
	
	\bibitem{E5-3-1} Sh.K. Pandey and J. Feng, {\it  Spin interaction and magnetism in cobaltate Kitaev candidate materials: An ab initio and model Hamiltonian approach}, Phys. Rev. B {\bf 106}, 174411 (2022), \doi{10.1103/PhysRevB.106.174411}.
	
	\bibitem{E5-3-2} W. Yao, Y. Zhao, Y. Qiu, C. Balz, J.R. Stewart, J.W. Lynn and Y. Li, {\it Magnetic ground state of the Kitaev $Na_2Co_2TeO_6$ spin liquid candidate}, Phys. Rev. Research {\bf 5}, L022045 (2023), \doi{10.1103/PhysRevResearch.5.L022045}.
	
	
	
	
	\bibitem{E5-4} J.G. Rau, E.K.H.  Lee and H.Y.  Kee, {\it  Generic Spin Model for the Honeycomb Iridates beyond the Kitaev Limit}, Phys. Rev. Lett. {\bf 112}, 077204 (2014), \doi{10.1103/PhysRevLett.112.077204}.
	
	
	\bibitem{E5-5} Z. Zhao, T.C. Yi, M. Xue and W.L. You, {\it Characterizing quantum criticality and steered coherence in the XY-Gamma chain}, Phys. Rev. A {\bf 105}, 063306 (2022), \doi{10.1103/PhysRevA.105.063306}.
	
	\bibitem{E5-6} Z.A. Liu, Y.L. Dong, N. Wu, Y. Wang and W.L. You, {\it  Quantum criticality and correlations in the Ising-Gamma chain}, Physica A {\bf 579}, 126122 (2021), \doi{10.1016/j.physa.2021.126122}.
	
	\bibitem{E5-7}Q. Luo, J. Zhao, X. Wang, and H.Y. Kee, {\it Unveiling the phase diagram of a bond-alternating spin-1/2 $K-\Gamma$ chain}, Phys. Rev. B {\bf 103}, 144423 (2021), \doi{10.1103/PhysRevB.103.144423}.
	
	\bibitem{E5-8} W. Yang, A. Nocera, C. Xu, A. Adhikary and I. Affleck, {\it Emergent  SU(2)$_1$ conformal symmetry in the spin-1/2 Kitaev-Gamma chain with a Dzyaloshinskii-Moriya interaction}, \doi{10.48550/arXiv.2204.13810}.
	
	\bibitem{E5-9-1} Sasan Kheiri, Hadi Cheraghi, Saeed Mahdavifar, and Nicholas Sedlmayr, {\it Information propagation in one-dimensional XY-$\Gamma$ chains}, Phys. Rev. B {\bf 109}, 134303 (2024), \doi{10.1103/PhysRevB.109.134303}.
	
	
		\bibitem{E5-12-0} P. Jordan and E. Wigner, {\it  Über das Paulische Äquivalenzverbot}, Z. Phys. {\bf 47}, 631 (1928), \doi{10.1007/BF01331938}.
	
	
	
	\bibitem{E5-12} S. Yamamoto and K. Funase, {\it Fermionic versus bosonic descriptions of one-dimensional spin-gapped antiferromagnets, Low Temp}. Phys. {\bf 31}, 740 (2005), \doi{10.1063/1.2008134
	}.
	\bibitem{E5-12-p1}
	 J. Abouie and S. Mahdavifar, {\it Signature of the Luttinger liquid phase in alternating Heisenberg spin-1/2 chains: Analytical and numerical approaches}, Phys. Rev. B {\bf 78}, 184437 (2008), \doi{10.1103/PhysRevB.78.184437}.
	
	
	
	\bibitem{E5-9} ] N. Avalishvili, G.I. Japaridze and G. L. Rossini, {\it  Long-range spin chirality dimer order in the Heisenberg chain with modulated Dzyaloshinskii-Moriya interactions}, Phys. Rev. B {\bf 99}, 205159 (2019), \doi{10.1103/PhysRevB.99.205159}.
	
	\bibitem{E5-10} F. K. Fumani, B. Beradze, S. Nemati, S. Mahdavifar, and G. I. Japaridze, {\it Quantum correlations in the spin-1/2 Heisenberg XXZ chain with modulated Dzyaloshinskii-Moriya interaction}, J. Magn. Magn. Mater. {\bf 518}, 167411 (2020), \doi{10.1016/j.jmmm.2020.167411}.
	
	\bibitem{E5-11} G. I. Japaridze, H. Cheraghi and S. Mahdavifar, {\it Magnetic phase diagram of a spin-1/2 XXZ chain with modulated Dzyaloshinskii-Moriya interaction}, Phys. Rev. E {\bf 104}, 014134 (2021), \doi{10.1103/PhysRevE.104.014134}.
	
	\bibitem{R1-1} A. Langari and S. Mahdavifar, {\it Gap exponent of the XXZ model in a transverse field}, Phys. Rev. B {\bf 73}, 054410 (2006), \doi{10.1103/PhysRevB.73.054410}.
	
	\bibitem{R1-2} S. Mahdavifar, {\it Scaling behaviour of the energy gap of spin-1/2 AF-Heisenberg chains in both uniform and staggered fields}, Eur. Phys. J. B. {\bf 55}, 371 (2007), \doi{10.1140/epjb/e2007-00068-8}.
	
	
	\bibitem{R1-4} M. Fujihala, M. Hagihala, K. Morita, N. Murai, A. Koda, H. Okabe, {\it and S. Mitsuda, Spin gap in the weakly interacting quantum spin chain antiferromagnet $KCuPO_4\cdot H_2O$}, Phys. Rev. B {\bf 107}, 054435 (2023), \doi{10.1103/PhysRevB.107.054435}.
	
	
	\bibitem{R1-5}  C.H.L. Quay, T.L. Hughes, J.A. Sulpizio, L.N. Pfeiffer, K.W. Baldwin, K.W. West, D. Goldhaber-Gordon  and R. de Piccioto, {\it  Observation of a one-dimensional spin-orbit gap in a quantum wire}, Nat. Phys. {\bf 6}, 336 (2010), \doi{10.1038/nphys1626}. 
	
	\bibitem{R1-6} R. Mao, Y.W. Dai, S.Y. Cho and H.Q. Zhou, {\it Quantum coherence and spin nematic to nematic quantum phase transitions in biquadratic spin-1 and spin-2 XY chains with rhombic single-ion anisotropy}, Phys. Rev. B {\bf 103}, 014446 (2021), \doi{10.1103/PhysRevB.103.014446}.
	
	\bibitem{R1-7} W. Yang, A. Nocera, E.S. Sørensen, H.Y. Kee and I. Affleck, {\it Classical spin order near the antiferromagnetic Kitaev point in the spin-1/2 Kitaev-Gamma chain}, Phys. Rev. B {\bf 103}, 054437 (2021), \doi{10.1103/PhysRevB.103.054437}.
	
	
	\bibitem{R1-8} S. Furukawa, M. Sato, S. Onoda and A. Furusaki, {\it Ground-state phase diagram of a spin-1/2 frustrated ferromagnetic XXZ chain: Haldane dimer phase and gapped/gapless chiral phases}, Phys. Rev. B {\bf 86}, 094417 (2012), \doi{10.1103/PhysRevB.86.094417}.
	
	
	
	
	


\end{thebibliography}
\end{document}